\documentclass[twocolumn,superscriptaddress,nofootinbib,floatfix,aps,prb,citeautoscript,longbibliography]{revtex4-2}
\usepackage[utf8]{inputenc}
\usepackage[T1]{fontenc}
\usepackage{lineno}

\usepackage{graphicx}
\usepackage{color}
\usepackage{array}
\usepackage{dcolumn}
\usepackage{bm}
\usepackage{multirow}
\usepackage[version=4]{mhchem}
\usepackage{cleveref}
\usepackage{tabularx}
\usepackage[usenames,dvipsnames]{xcolor} 
\usepackage{siunitx}
\usepackage{comment}

\newcommand{\Tc}{\ensuremath{T_\textrm{c}}}

\newcommand{\Tcad}{\ensuremath{T_\textrm{c}^\text{Allen-Dynes}}}
\newcommand{\Tce}{\ensuremath{T_\textrm{c}^\text{Eliashberg}}}
\newcommand{\olog}{\ensuremath{\omega_\text{log}}}
\newcommand{\dosef}{\ensuremath{\text{DOS}({\text{E}_\text{F}})}}
\newcommand{\afo}{\ensuremath{\alpha^2F(\omega)}}

\newcommand{\etal}{{\textit et al.}}

\begin{document}

\newcommand{\bochum}{Research Center Future Energy Materials and Systems of the University Alliance Ruhr, Faculty of Mechanical Engineering, Ruhr University Bochum, Universitätsstraße 150, D-44801 Bochum, Germany}
\newcommand{\coimbra}{CFisUC, Department of Physics, University of Coimbra, Rua Larga, 3004-516 Coimbra, Portugal}
\newcommand{\mpi}{Max-Planck-Institut f\"ur Mikrostrukturphysik, Weinberg 2, D-06120 Halle, Germany}

\author{Tiago F. T. Cerqueira}
\affiliation{\coimbra}
\author{Antonio Sanna}
\affiliation{\mpi}
\author{Miguel A. L. Marques} 
\email{miguel.marques@rub.de}
\affiliation{\bochum} 

\date{\today}

\title{Sampling the Whole Materials Space for Conventional Superconducting Materials}

\begin{abstract}
We perform a large scale study of conventional superconducting materials using a machine-learning accelerated high-throughput workflow. We start by creating a comprehensive dataset of around 7000 electron-phonon calculations performed with reasonable convergence parameters. This dataset is then used to train a robust machine learning model capable of predicting the electron-phonon and superconducting properties based on structural, compositional, and electronic ground-state properties. Using this machine, we evaluate the transition temperature (\Tc) of approximately 200\,000 metallic compounds, all of which on the convex hull of thermodynamic stability (or close to it) to maximize the probability of synthesizability. Compounds predicted to have \Tc\ values exceeding 5~K are further validated using density-functional perturbation theory. As a result, we identify 545 compounds with \Tc\ values surpassing 10~K, encompassing a variety of crystal structures and chemical compositions. 
This work is complemented with a detailed examination of several interesting materials, including nitrides, hydrides, and intermetallic compounds. Particularly noteworthy is \ce{LiMoN2}, which we predict to be superconducting in the stoichiometric trigonal phase, with a \Tc\ exceeding 38~K. \ce{LiMoN2} has been previously synthesized in this phase, further heightening its potential for practical applications.
\end{abstract}

\maketitle

\section{Introduction}

Superconducting materials are used in a growing number of applications crucial to technological, societal and economical development. These include low magnetic field systems like cables, transformers and motors, qubits for quantum computers and, more significantly, high field devices for medical diagnostic, accelerators in particle physics and plasma confinement for fusion technology~\cite{Hull_AppliedSCchapter4_SCMagnets2015,Yao_SCmaterialsChallengesOpportunitiesApplication2021}.  
Hundreds of superconducting material families have been discovered, among which the copper-oxides~\cite{Bednorz_Muller86} with critical temperatures (\Tc) exceeding 100~K, MgB$_{\text 2}$ with \Tc=39~K~\cite{Nagamatsu_MgB2_Nature2021}, the ferro-pnictides reaching a \Tc\ of 55~K~\cite{Kamihara2006,Kamihara2008}, and more recently the high pressure hydrides~\cite{DrozdovEremets_SH3_Nature2015,Hemley_LaH_PRL2019,Eremets_LaH_Nature2019,FloresLivas_Perspective_PhysReports2020}, with critical temperatures close to room values.

It is then, at a first sight, surprising that some of the largest experiments in physics, like the Large-Hadron Collider at CERN or the International Thermonuclear Experimental Reactor (our biggest hope of developing a source of clean and nearly infinite energy), rely on superconductors
---  Nb--Ti alloys and Nb$_{\text 3}$Sn --- that were developed half a century ago and were already in use in the 1970s~\cite{Powell_LargeScaleApplications_BookChapter1974}. 
The reason is that, despite their remarkable superconducting properties, most high-Tc materials are extremely brittle and therefore poorly suited to make the many kilometers of wire or tape required~\cite{Yao_SCmaterialsChallengesOpportunitiesApplication2021,Gurevich_ToUseOrNotToUseCoolSuperconductors_NatPhys2011}. In addition, cuprates suffer from the grain boundary problem which makes the construction of long stable cables complex and expensive.
In contrast, Nb--Ti forms ductile alloys, which are well suitable for fabricating Cu-stabilized multifilamentary conductors, but have a low critical temperature. However, Nb is an expensive chemical element of limited availability.

In view of this situation, it is urgent to discover new industry-friendly materials that may resolve, or at least alleviate, the current dependence of the industry on Nb-based superconductors. Over the last years several high-profile projects were aimed to discover new superconductors, including a large-scale 4-year proposal financed by the Japanese government~\cite{Hosono_NewSuperconductorsFe_2015}. The overwhelming majority of the candidate systems turned out to be non-superconducting, and only 100 new superconducting materials where found, mostly iron-based or analogous to them. This gives a clear indication of the low time- and cost-effectiveness of the traditional experimental trial-and-error search for superconductors.

A targeted approach where synthesis is only attempted following a precise theoretical prediction could be more efficient. This is possible for conventional superconductors, that are well understood, and for which predictive quantitative theories exist~\cite{BoeriBacheletViewpoint2019,marsiglioEliashbergTheoryShort2020,Lueders_SCDFT_PRB2005,Marques_SCDFT_PRB2005,SPG_EliashbergSCDFT_PRL2020}. Unfortunately search has been mostly focused on selected chemical compositions and small families of compounds~\cite{FloresLivas_Perspective_PhysReports2020,BoeriRoadmap2022,Kolmogorov_FeB_PRL2010,Zurek_CB_JACS2023,Boeri_Boron_PRB2020,Ma_dopedhydrides_PRL2019}, while very few large scale studies of  superconductors can be found in the literature. In 2022, Hoffmann \etal\ studied the whole family of antiperovskites~\cite{Hoffmann2022}, performing electron-phonon calculations for more than 400 compounds. This was followed by a similar work on full-Heusler materials comprising more than 1000 candidate systems, and discovering a total of 8 hypothetical materials with critical temperatures above 10~K~\cite{ourHeusler}. In spite of relatively large number of chemical compositions studied, these works were however strongly limited in terms of the possible crystal structures. 

In a seminal work, Choudhary and Garrity~\cite{Choudhary2022} overcame this limitation by devising a multi-step, machine-learning accelerated workflow for discovering conventional superconductors. They conducted electron-phonon coupling calculations for 1058 compounds with varying structures and chemical compositions, leading to the identification of 105 dynamically stable materials with \Tc\ above 5~K, but only 17 with \Tc\ above 10~K and having distances to the hull of stability smaller than 50~meV/atom. We strongly believe that this approach is the right path to follow, because it targets the complete space of possible materials. However the work of Choudhary and Garrity suffered from two limiting aspects: the use of a 2x2x2, severely underconverged $q$-point mesh, and the small dataset size which was insufficient to construct a reliable machine learning model.

In this work we take a step further by creating a \textit{large and accurate} dataset of calculated electron-phonon and superconducting properties that is used to train a \textit{robust} machine-learning model for predicting the transition temperature of conventional superconductors. This model is then used to sieve through almost 200\,000 compounds to uncover a large number of conventional superconductors with high-\Tc.

The remainder of this article is structured as follows:  Sec.~\ref{sec:strategy} explains our general strategy and workflow. The training dataset is described and analyzed in Sec.~\ref{sec:DBI}, providing insights about the properties of conventional superconductors and their statistical distribution. 
Sec.~\ref{sec:entries} is devoted to our predictions, presenting the most interesting compounds we have found and providing a full characterization of a few noteworthy materials. Finally we present our conclusions in Sec.~\ref{sec:conclusion}.

\section{Strategy and Workflow}
\label{sec:strategy}

\begin{figure}
  \centering
    \includegraphics[width=\columnwidth]{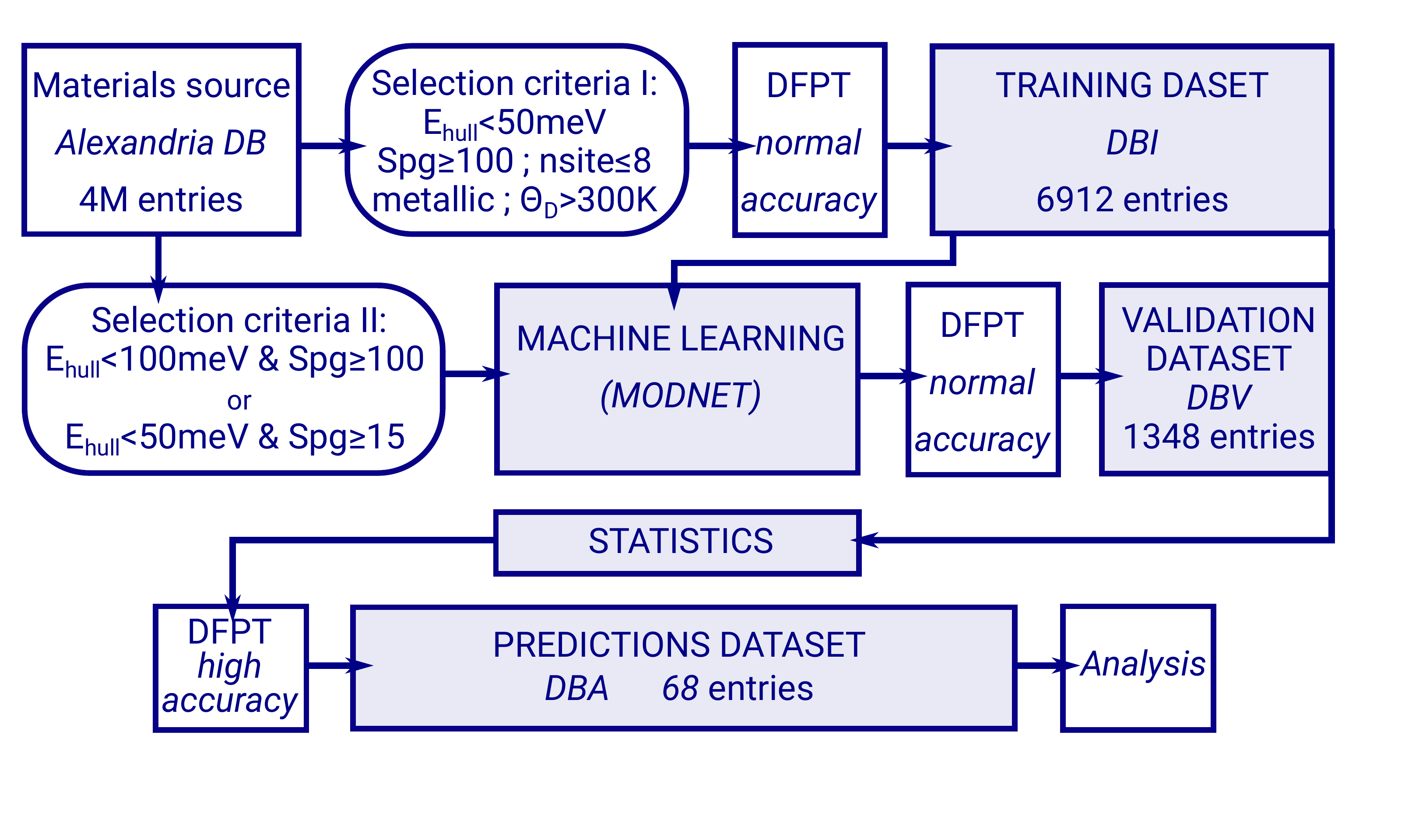}
  \caption{Sketch of our search workflow. We start from a large database of crystal structures and Density Functional Theory calculations. From this we select and compute (via DFPT with \textit{normal} accuracy) a training database (DBI) whose superconducting properties are computed from first principles. DBI is used for training a machine learning model. The machine is used to identify promising materials which are simulated and included into
  the validation database (VDB). Among all computed materials (DBA=DBI+DBV) we perform a statistical analysis and select the best systems with respect to their superconducting properties. This constitutes our prediction set, for which we perform a detailed analysis.}
  \label{fig:workflow}
\end{figure}

Our search workflow is summarized in Fig.~\ref{fig:workflow} and commences with the creation of a large dataset of electron-phonon calculations. The selection of materials for this dataset is a critical step that requires careful consideration. On the one hand, the choice should aim to minimize possible biases in the representation of chemical elements and crystal structures, ensuring a diverse and representative sample. On the other hand, the computational effort required for electron-phonon calculations increases steeply with the number of atoms in the unit cell, necessitating thorough assessment of computational feasibility.

We start with the stable or nearly stable compounds (below $50$~meV/atom from the convex hull of thermodynamic stability) from the \textsc{Alexandria} database of Refs.~\onlinecite{schmidtCrystalGraphAttention2021,10.1002/adma.202210788}. 
We discarded semiconductors, insulators and any material with nonzero magnetic moments as these should not host a stable superconducting state with large critical temperature. Also semi-metallic systems and systems with very low density of states at the Fermi level (\dosef) are not relevant and were filtered out. 

In addition to these, somehow straightforward, selection criteria, we added a further filter by including only materials with an estimated Debye temperature ($\Theta_\text{D}$) above a cutoff of $300$~K. Although there is no simple proportionality between the Debye temperature and \Tc, it is a fact that the best superconductors have relatively large phonon frequencies. As the \textsc{Alexandria} dataset does not include the values for the Debye temperature, these were estimated using a crystal-graph neural network (see Sec.~\ref{sec:methMachineLearning}). We note that the cutoffs in \dosef\ and $\Theta_\text{D}$ were already used in Refs.~\onlinecite{Choudhary2022,10.1002/adma.202210788}.

For numerical efficiency reasons, we also limited the training set to compounds with $\le8$ atoms in the primitive unit cell, and space group number $\ge100$ (including most tetragonal, and all trigonal, hexagonal, and cubic lattices, but excluding the orthorhombic, monoclinic and triclinic systems).

Subsequently, the entries were arranged based on Pareto fronts within the \dosef\ and $\Theta_\text{D}$ parameter space, and the calculation of the electron-phonon interaction was performed with density-functional perturbation theory~\cite{PhysRevLett.58.1861,Giannozzi2017,Giannozzi2009} (DFPT) in that specified sequence (refer to Sec.~\ref{sec:methods}). To increase the diversity of the dataset, we added entries with few atoms in the unit cell ($\le 5$) from higher Pareto fronts. Finally, we also included the 50 materials already calculated by some of us in Ref.~\onlinecite{10.1002/adma.202210788}, and the datasets of Heuslers from Ref.~\onlinecite{ourHeusler} and of inverted perovskites from Ref.~\onlinecite{Hoffmann2022}. These later two were originally calculated with the LDA functional~\cite{PhysRevB.45.13244} and with $k$ and $q$-point sets specifically chosen for those crystal structures. In order to make the data compatible with our current approach, the ground-state and the electron-phonon coupling constants were recalculated with the parameters presented in Sec.~\ref{sec:methods}. Note that a table containing a summary of all our DFPT calculations can be found in the Supplemental Information (SI).

Our training dataset (DBI) contains results for 6912 dynamically stable materials, of which 2239 have a \Tcad\ (\Tc\ estimated with the Allen-Dynes formula~\cite{AllenDynes_PRB1975}) larger than 1~K, 678 larger than 5~K, and 214 larger than 10~K. This dataset, analyzed in detail in Sec.~\ref{sec:DBI}, was used to train machine-learning models for predicting \olog, $\lambda$, and the superconducting transition temperature. The models considered structural, compositional, and ground-state features, and we employed \textsc{modnet}~\cite{DeBreuck2021} and \textsc{alignn}~\cite{Choudhary2021} using various inputs, outputs, and training strategies. Our most successful crystal-graph network was a \textsc{modnet} model using as input features the structure and composition of the compound together with \dosef\ and $\Theta_\text{D}$ (see \ref{sec:methMachineLearning}). The output was the vector composed of \olog, $\lambda$, and \Tc\ and the cost function was the linear combination (with equal weights) of the error in these three quantities.

We applied this machine to predict the transition temperature of three further datasets, that we refer together as our validation dataset (DBV). As before, we excluded magnetic and semiconducting compounds. In all cases, we selected all compounds predicted to have a \Tc\ larger than 5~K and performed validation runs with DFPT of the electron-phonon interaction. 

(i)~First we selected metals with a maximum of 8 atoms in the primitive unit cell, including trigonal, hexagonal, and cubic systems, present in the database~\cite{schmidtCrystalGraphAttention2021,10.1002/adma.202210788} within 50~meV/atom from the convex hull (amounting to a material space of 108771 entries). We obtained with DFPT 514 dynamically stable compounds, of which 209 had a transition temperature above 5~K. This resulted in a success rate of 40\%.

(ii)~In this case we relaxed the thermodynamic stability constrain, and allowed for compounds with a maximum of 5 atoms in the primitive unit cell, including trigonal, hexagonal, and cubic systems, present in the database~\cite{schmidtCrystalGraphAttention2021,10.1002/adma.202210788} between 50 and 100~meV/atom from the convex hull. This amounted to a material space of 65288 compounds. We found 721 dynamically stable phases, from which 549 had a transition temperature above 5~K. This resulted in a success rate of 76\%.

(iii)~To investigate lower symmetry compounds, we selected orthorhombic and tetragonal compounds with a maximum of 5 atoms in the primitive unit cell present in the database~\cite{schmidtCrystalGraphAttention2021,10.1002/adma.202210788} within 50~meV/atom from the convex hull. This amounted to a material space of 17469 compounds. We found 114 dynamically stable phases, of which 72 had a transition temperature above 5~K. This resulted in a success rate of 63\%.

Together, we searched a material space of 191528 compounds with our machine learning model, with an average success rate to find compounds with $\Tc>5$~K of 65\%. We note that in the initial dataset DBI only around 10\% of the compounds were found to have $\Tc>5$~K, proving the efficiency of our approach.

We estimated the precision on \Tc\ of our machine in this validation set with a mean of $\Tc=8.07$~K, obtaining a mean absolute error of 2.74~K. To put this figure into perspective, we can consider two hypothetical machines: (i)~A machine that predicts $\Tc=0$~K for all entries would yield a mean absolute error of 8.07~K, which corresponds to the mean value of the test set; (ii)~Alternatively, a machine that predicts $\Tc$ to be the mean of the values in the training set (1.63~K) would result in a mean absolute error of 5.56~K. For comparison, the machine in Ref.~\onlinecite{Choudhary2022} had an error of 1.84~K for a mean value of the training set of 2.72~K. We attribute this notable improvement to the substantial expansion of the training set, which was around ten times larger than the dataset used in Ref.~\onlinecite{Choudhary2022}. As a result, our model can now make \textit{robust} predictions of superconducting compounds solely based on their structural, compositional, and ground-state properties.

The described workflow, consisting of multiple accuracy steps, may appear complicated. However, given the inherent complexity involved in the search for superconductors, it appears to us as the most viable approach for accelerating the discovery process.

\section{Analysis of the dataset}
\label{sec:DBI}

In this section, our focus lies on the training dataset DBI. The quality of our \textit{ab initio} simulations enables us to conduct a meaningful statistical analysis of this set.
Our primary objective is to identify the key properties associated with high \Tc. Most importantly, we aim at estimating the rarity of finding conventional superconductors with a given critical temperature. 
Here, we do not include the materials found by machine-learning in order to reduce the bias in the statistical distributions. The DBI set may still possess some bias due to our initial selection process, with the purpose of excluding non-superconducting systems and materials with excessive computational cost. However these are relatively minor issues and, we believe, that the dataset still holds valuable statistical information.  

\subsection{Chemical space}

\begin{figure}
  \centering
  (a)\\
  \includegraphics[height=6.cm]{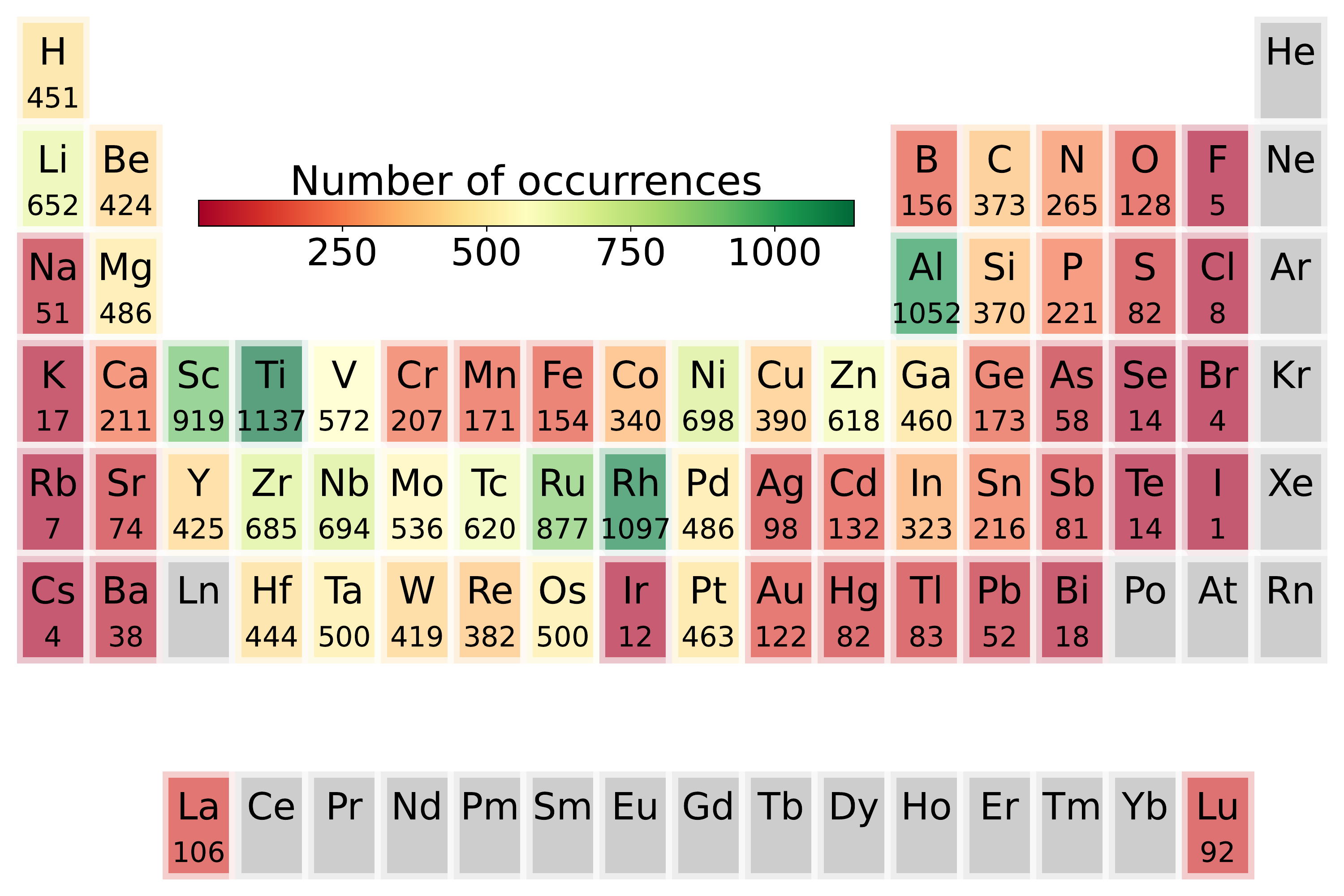}\\[5mm]
  (b)\\
  \includegraphics[height=6.cm]{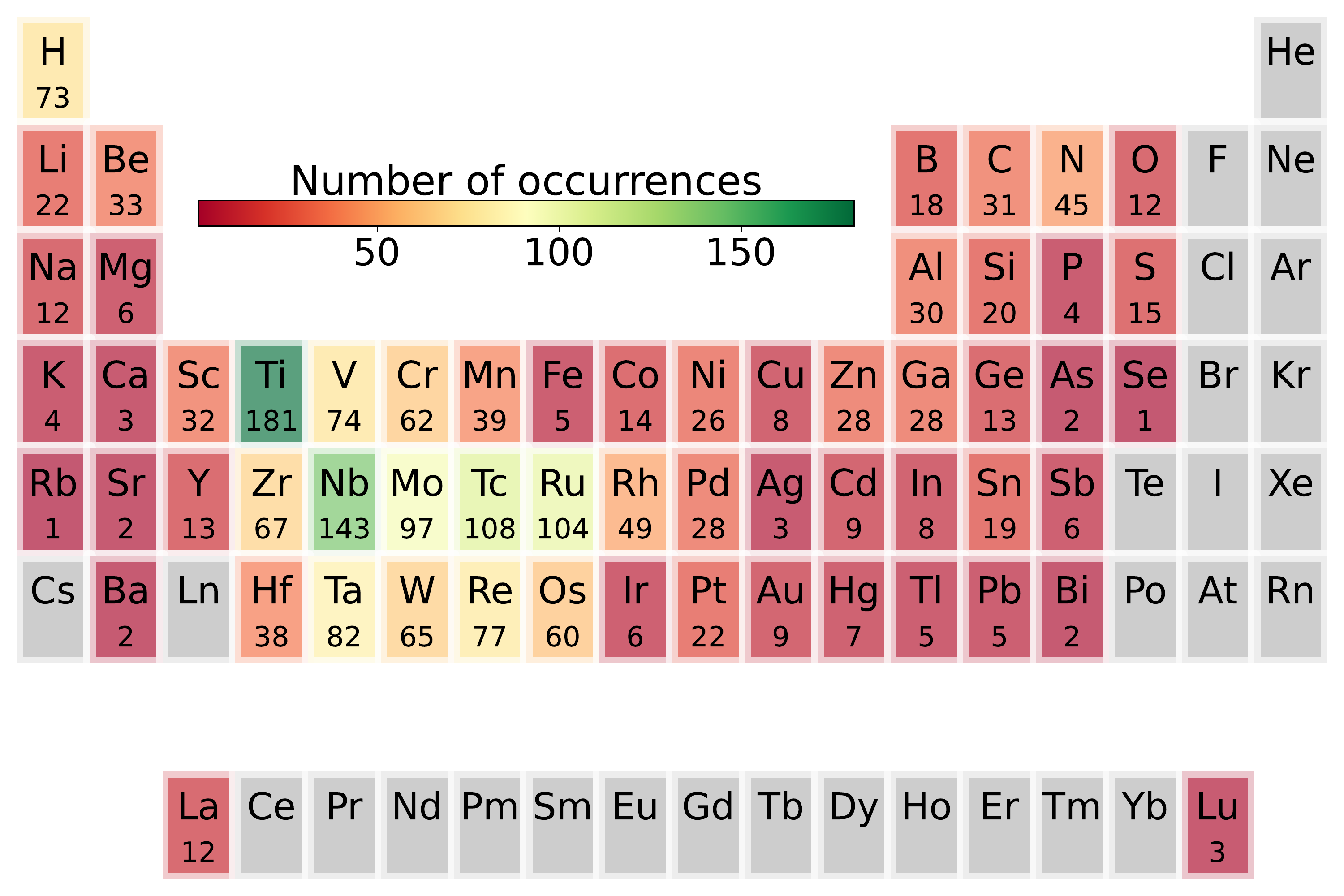}
  \caption{Periodic table with the number of occurrences of each chemical for (a)~all compounds in DBI and (b)~compounds in DBI with a \Tcad\ larger than 5~K.}
  \label{fig:periodictable}
\end{figure}

In \ref{fig:periodictable}(a) we plot the distribution of chemical elements in our DBI. The boxes in gray represent either chemical elements for which we did not have pseudopotentials available or the rare gases that do not form metallic compounds at ambient conditions. We would also like to note that we only have few compounds with Ir in DBI due to the problems with its pseudopotential mentioned in Sec.~\ref{sec:methods}. In general, the number of occurrences decreases with the period. The exception are the $3d$ elements Cr, Mn, Fe, etc. that have the tendency to form magnetic compounds that were removed from the dataset. Non-metals are also much less represented than metals, as most systems in DBI are intermetallic compounds. Also we note the scarcity of compounds with heavy alkali and alkali earth elements.

If we now look at the distribution of the chemical elements present in compounds having $\Tcad > 5$~K we obtain a completely different picture. As expected many superconducting compounds contain hydrogen. 
However the largest fraction of superconducting compounds include early transition metals with a peak appearance of Nb and Ti compounds. 
Moreover, the ideal group for superconductivity appears to be group VI, with 30\% of Cr-, 18\% of Mo-, and 16\% of W-compounds having $\Tcad > 5$~K, followed by group VII (Mn, Tc, Re) and group V (V, Nb, Ta).  However, the latter group has the advantage of being less magnetic so superconductivity does not have to compete with disrupting effects as spin fluctuations (which we do not consider in this work). 
The occurrence of noble metals and non-metals is very much reduced. However, as we will see in the following, while not statistically prevalent, we do find compounds with extreme values of \Tc\ including these elements. Another interesting fact relates to the second row of the periodic table: we find an increase of high-temperature superconducting systems going from B to C, arriving at a maximum for N (17\% of its compounds having $\Tcad > 5$~K), and then decreasing for O and F. This behavior is not mirrored in the third row where the ideal element for superconductivity is sulphur (19\% with $\Tcad > 5$~K). This indicates that nitrogen and sulphur might be very favorable targets for the design of high-\Tc\ conventional superconductors, better than the well-known B and C.

\subsection{Superconducting properties}
\label{sec:distributions}

\begin{figure}
  \centering
  (a)\\
  \includegraphics[height=5.cm]{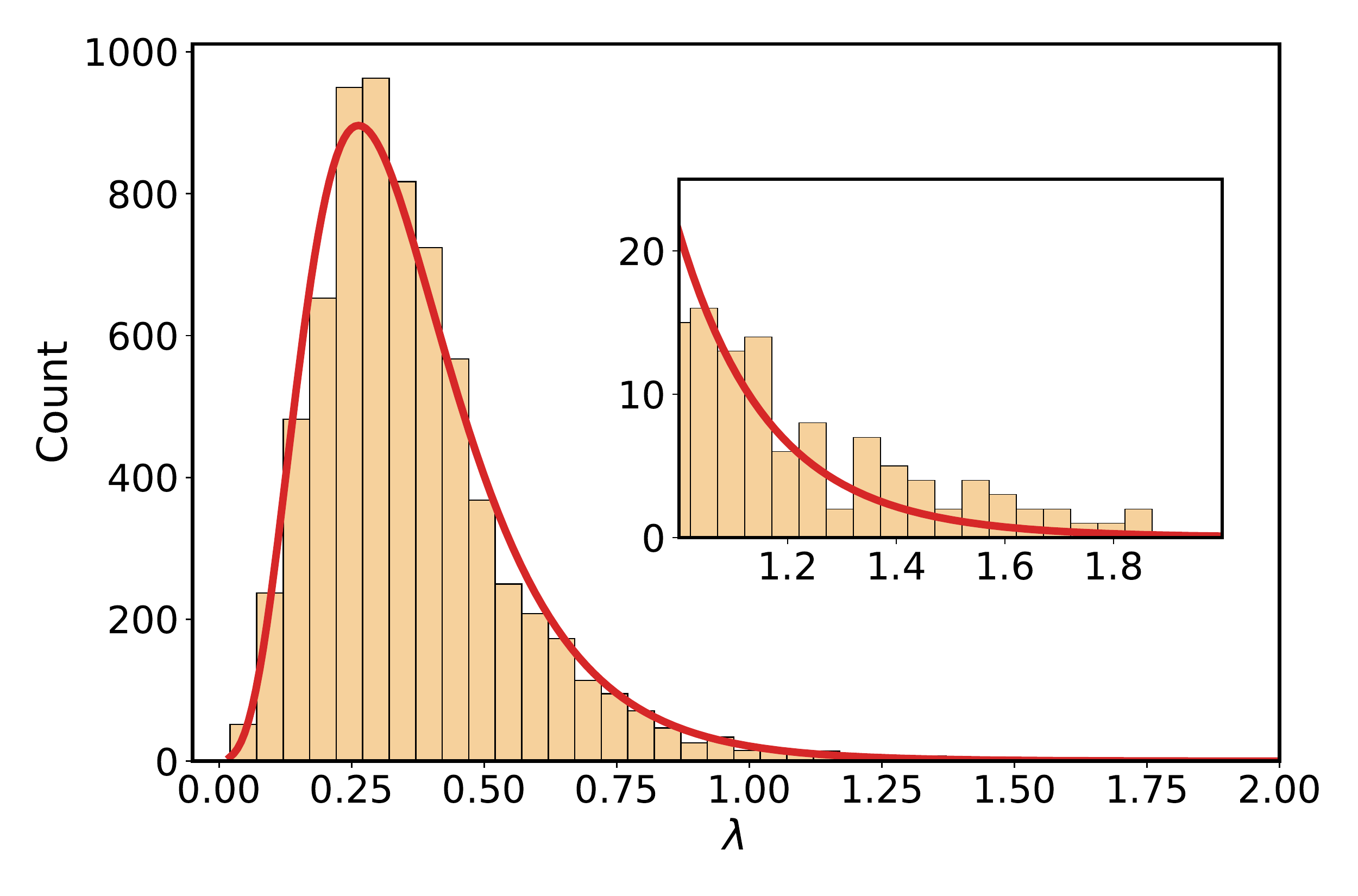}\\
  (b)\\
  \includegraphics[height=5.cm]{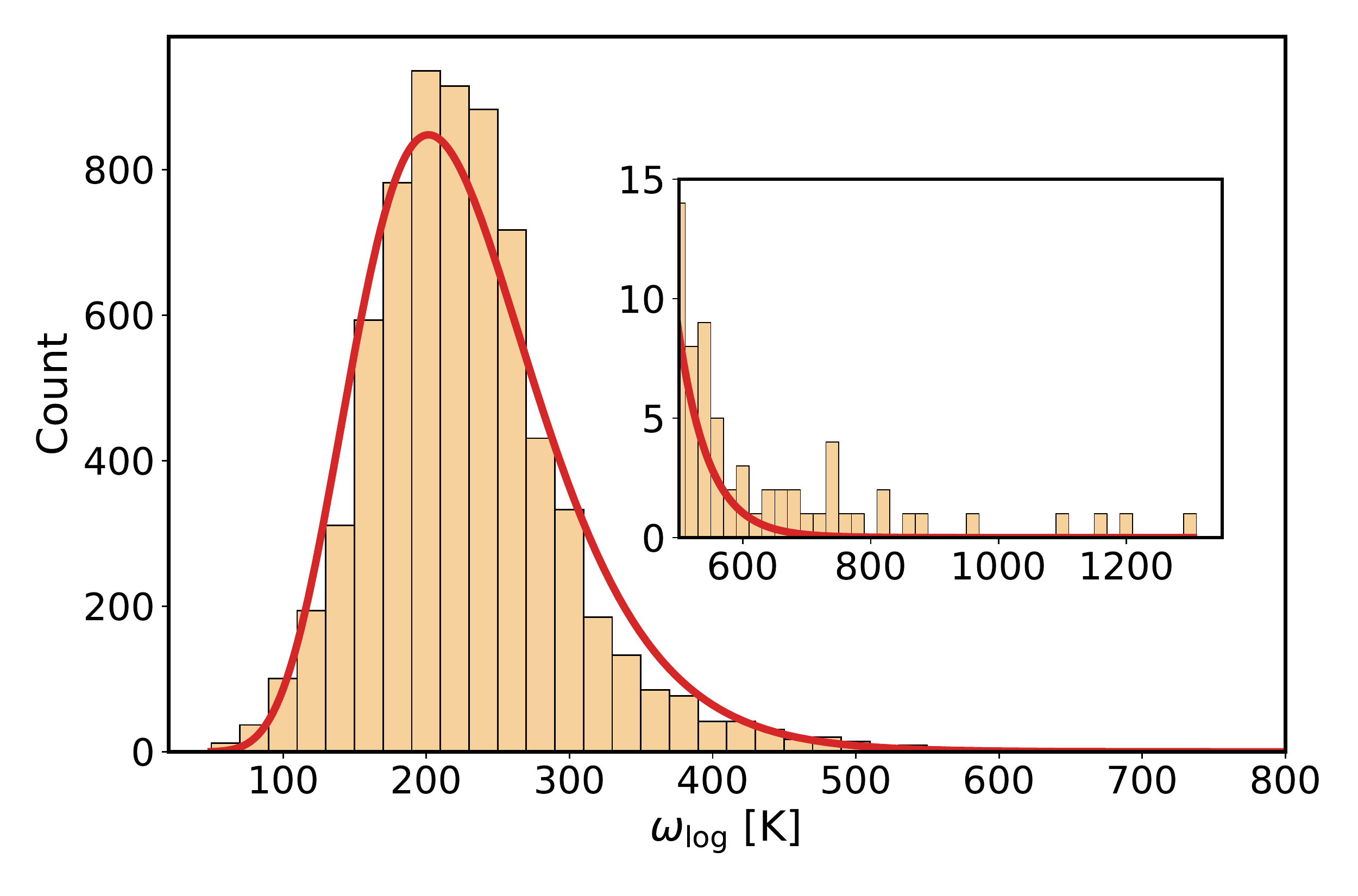}\\
  (c)\\
  \includegraphics[height=5.cm]{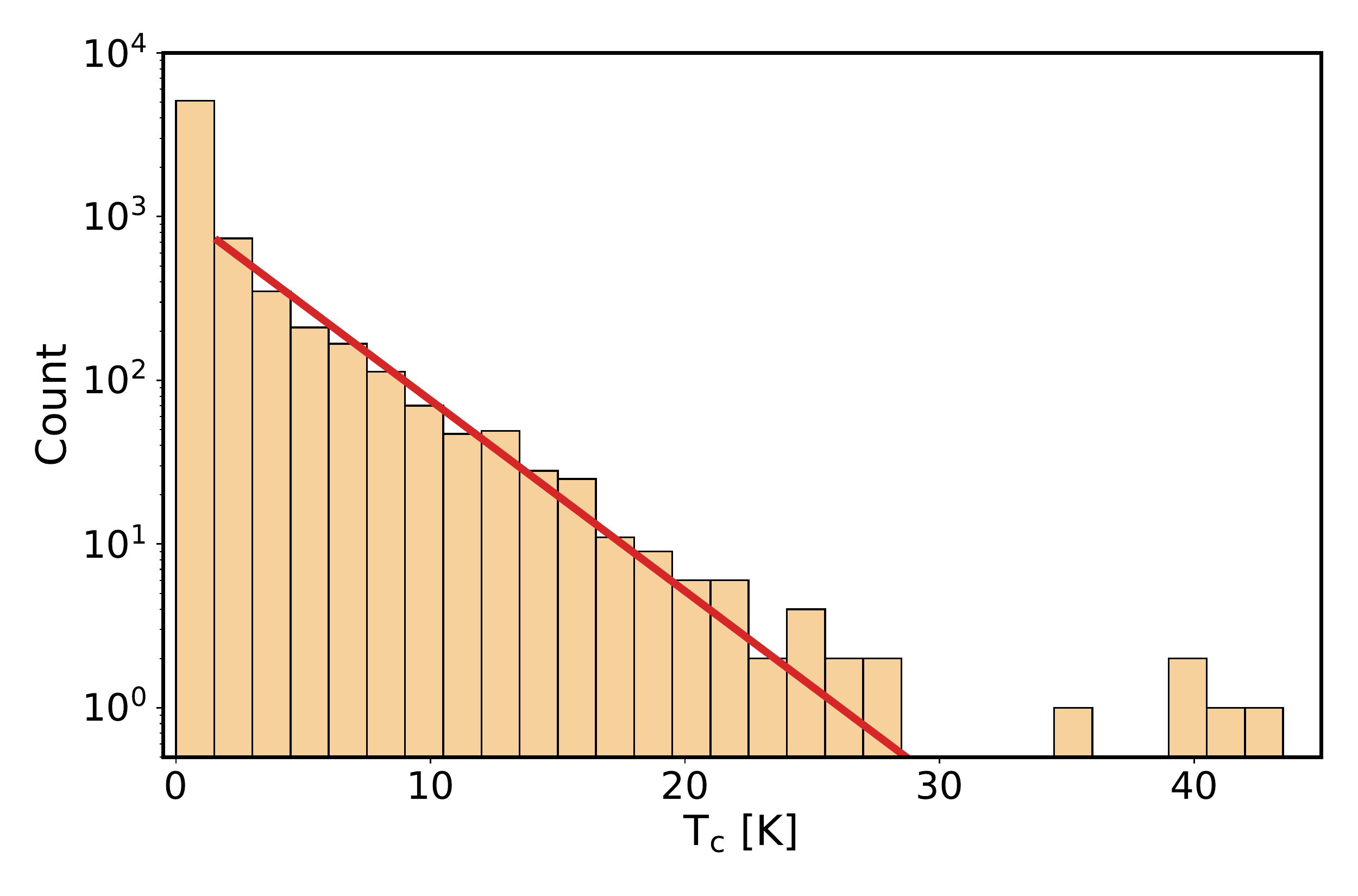}

  \caption{(a)~Histogram of the electron-phonon coupling constant $\lambda$ with bins of width 0.05. The red curve is a fit to a lognormal distribution $f(x, s) = \frac{1}{s d x \sqrt{2\pi}}\exp\left(-\frac{\log^2(y)}{2s^2}\right)$, with $y = (x - \bar x) / d$, yielding the parameters $s=0.433$, $\bar x = -0.0618$, and $d=0.392$; (b)~Histogram of the logarithmic averaged phonon frequency $\olog$ with bins of width 20~K; The red curve is a fit to a lognormal distribution with parameters $s=0.276$, $\bar x = -26.3$, and $d=246$; (c)~Histogram of the values of the transition temperature \Tc\ with bins of width 1.5~K. Note the logarithmic scale on the $y$-axis. The red curve is a fit, performed for entries with $\Tc > 1.5$~K, of an exponential distribution $\exp\left(-y\right)/d$, yielding $\bar x = 1.50$ and $d = 3.72$. The fits were performed with \textsc{scipy}~\cite{2020SciPy-NMeth}.}
  \label{fig:histograms}
\end{figure}

In Fig.~\ref{fig:histograms} we plot histograms of the electron-phonon and superconducting properties as calculated in our DBI dataset. Due to the large amount of compounds, the histograms are very smooth, in contrast to results for Heusler or anti-perovskite compounds only~\cite{Hoffmann2022,ourHeusler}.

In panel (a) we see that the distribution of values of $\lambda$ is, as expected, very asymmetric, with a maximum value at around 0.25 and with a fat tail that extends well beyond $\lambda=1$ (see inset). The mean value of the distribution $\lambda_\text{mean}=0.37$) is in very good agreement with the mean value found for Heuslers ($\lambda_\text{mean}=0.30$) and anti-perovskites ($\lambda_\text{mean}=0.36$). Surprisingly, the histogram can be fit by a lognormal distribution, allowing us to easily calculate probabilities. Note that, strictly speaking, these are conditional probabilities reflecting any residual bias present in the choice of materials contained in the dataset of Ref.~\cite{schmidtCrystalGraphAttention2021,10.1002/adma.202210788}, and of our choices detailed in Sec.~\ref{sec:strategy}. With this in mind, we find that the probability to find a material with $\lambda$ greater than 0.5 is 19.6\%, greater than 0.75 is 4.5\%, and greater than 1.0 is 1.0\%. We note that if we zoom the plot in the region of the tail, we find many more compounds with large values of $\lambda$ than expected from the distribution. However, we can not use this fact as an evidence for the existence of a fat, Pareto tail, as many of these large values of $\lambda$ come from false positive materials with soft modes.

The histogram of \olog\ is shown in Fig.~\ref{fig:histograms}(b). As expected by the distribution of atomic masses, almost all materials exhibit \olog\ between around 100 and 400~K, with a maximum of around 200~K. The mean of the distribution is 229~K, again comparing well to the one of Heusler (190~K) and anti-perovskite (234~K) compounds. Not surprisingly, the rare cases of very large \olog\ shown in the inset of Fig.~\ref{fig:histograms}(b) are all compounds with light atoms like H. This distribution can also be fitted by a lognormal curve, although the quality of the fit is somewhat inferior to the one of $\lambda$. This may be related to the cutoff in $\Theta_\text{D}$ used to select the materials, that could lead to a artificial decrease of materials with small \olog. 

Finally, in panel (c) of Fig.~\ref{fig:histograms} we depict the distribution of values of \Tc\ calculated from the Allen-Dynes formula. We can observe that the large majority of compounds is not superconducting, or exhibits a very small ($<1.5$~K) transition temperature. For higher \Tc\ the number of compounds decreases exponentially until around 20--25~K, but a few outliers can be found with higher transition temperatures. Unfortunately, most of these outliers turn out to be false positives with soft phonon modes. These results are very much in line with the old assumption that conventional superconductivity is limited to about 30~K. In fact, while nothing prevents phononic superconductors to reach much higher critical temperatures, these cases appear to be extremely rare, at least within the restriction of being at ambient pressure, excluding anisotropy effects and considering only systems that are stable or close to thermodynamic stability.

The exact probabilities of higher critical temperatures are difficult to estimate. In our SBI set only two exceed 40~K (with our Eliashberg estimation) which leads to a probability of 0.03\%. Using the lognorm fit of the electron phonon coupling and of the phonon frequencies, and assuming the values of the three parameters of the Allen-Dynes equation to be uncorrelated, the Allen-Dynes formula predicts that the decrease of probability with \Tc\ is almost exponential. However the lognorm fit is not very precise at extreme values (as shown by the insets in Fig.~\ref{fig:histograms}) and extreme values of \Tc\ do arise from these deviations from the exponential tail. 

On the other hand we find a significant number of superconductors in the 20 to 30~K range of \Tc, in the order of 0.2\% to 0.5\% of the materials in DSI. 
Superconductors within this range of \Tc\ may not represent a technological game changer but could still be valuable for specific applications if superconductivity is accompanied by other desirable properties, like abundance of the atomic species, ductility and high critical fields and currents.

\section{Extreme Compounds}

\label{sec:entries}

\begin{table}[htb]
   \caption{Calculated superconducting properties with the accurate settings.  We present the chemical formula, the space group number (Spg), the number of atoms in the primitive unit cell (NSites), the distance to the convex hull $E_\text{hull}$ (calculated as in Ref.~\onlinecite{dataset}, in meV/atom), \olog\ (in K), $\lambda$, and the transition temperature calculated from the solution of the isotropic Eliashberg equation (\Tce\ in K). The transition temperature was obtained with $\mu^* = 0.10$.}
   \label{tab:accuratea}
   \begin{tabular*}{\linewidth}{@{\extracolsep{\fill}}rrrrrrr}
   Formula & Spg & NSites & E$_\text{hull}$ & \olog & $\lambda$ & \Tce \\
   \hline \\[-2mm]
\ce{LiMoN2} & 160 & 4 & 8 & 268.6 & 1.777 & 46.3 \\
\ce{LiPdH2} & 166 & 4 & 101 & 531.8 & 0.981 & 43.2 \\
\ce{HPd} & 225 & 2 & 35 & 342.5 & 1.239 & 37.4 \\
\ce{LiTcN2} & 122 & 8 & 0 & 241.4 & 1.325 & 28.4 \\
\ce{NaTcN2} & 122 & 8 & 0 & 260.7 & 1.211 & 27.4 \\
\ce{ZrH3} & 139 & 4 & 94 & 363.4 & 0.967 & 24.9 \\
\ce{V} & 229 & 1 & 5 & 204.6 & 1.329 & 24.9 \\
\ce{Nb3Zn} & 223 & 8 & 82 & 113.1 & 2.390 & 24.7 \\
\ce{Cr4ReW} & 216 & 6 & 40 & 189.9 & 1.414 & 24.3 \\
\ce{TiV2} & 139 & 3 & 41 & 213.6 & 1.258 & 24.0 \\
\ce{TiV} & 129 & 4 & 46 & 190.9 & 1.369 & 24.0 \\
\ce{LiMoN2} & 122 & 8 & 56 & 292.0 & 0.990 & 22.4 \\
\ce{TiNb3} & 123 & 4 & 32 & 150.3 & 1.556 & 21.9 \\
\ce{ZrTc2} & 227 & 6 & 29 & 149.9 & 1.443 & 20.2 \\
\ce{Ti} & 225 & 1 & 62 & 204.9 & 1.117 & 19.7 \\
\ce{LiVRu2} & 225 & 4 & 39 & 149.4 & 1.419 & 19.3 \\
\ce{Cr3Os} & 223 & 8 & 21 & 239.7 & 0.987 & 19.1 \\
\ce{MgB2} & 191 & 3 & 0 & 725.1 & 0.594 & 18.9 \\
\ce{NbC} & 225 & 2 & 47 & 330.3 & 0.835 & 18.5 \\
\ce{TiNbV4} & 216 & 6 & 36 & 148.3 & 1.275 & 17.7 \\
\ce{Cr2Re} & 164 & 3 & 41 & 252.2 & 0.895 & 17.1 \\
\ce{Nb} & 229 & 1 & 29 & 147.1 & 1.248 & 17.0 \\
\ce{ScMoC2} & 166 & 4 & 49 & 288.0 & 0.864 & 16.9 \\
\ce{MoH} & 225 & 2 & 0 & 241.4 & 0.931 & 16.8 \\
\ce{KB6} & 221 & 7 & 0 & 803.3 & 0.559 & 16.8 \\
\ce{Ti3Te} & 223 & 8 & 56 & 186.8 & 1.041 & 16.3 \\
\ce{V6CoSi} & 200 & 8 & 23 & 262.9 & 0.849 & 16.2 \\
\ce{RbB6} & 221 & 7 & 28 & 811.2 & 0.548 & 16.0 \\
\ce{TaB2} & 191 & 3 & 0 & 297.0 & 0.816 & 15.4 \\
\ce{CrH} & 225 & 2 & 0 & 310.6 & 0.767 & 15.3 \\
\ce{Zr} & 225 & 1 & 40 & 132.9 & 1.236 & 14.7 \\
\ce{Ti2H} & 166 & 3 & 45 & 171.1 & 1.005 & 14.1 \\
\ce{Ti2W} & 164 & 3 & 22 & 196.5 & 0.898 & 13.4 \\
\ce{ZrN} & 225 & 2 & 0 & 377.2 & 0.667 & 13.3 \\
\ce{TaNb} & 221 & 2 & 6 & 162.1 & 1.001 & 13.3 \\
\ce{Ti2Tc} & 139 & 3 & 9 & 211.5 & 0.831 & 12.6 \\
\ce{KCdH3} & 221 & 5 & 41 & 686.4 & 0.533 & 12.3 \\
\ce{Tc} & 194 & 4 & 10 & 213.7 & 0.808 & 12.2 \\
\ce{Be4NbRh} & 216 & 6 & 47 & 311.5 & 0.701 & 12.2 \\
\ce{LaRuH2N} & 123 & 5 & 16 & 337.0 & 0.690 & 12.0 \\
\ce{RuO2} & 136 & 6 & 0 & 444.4 & 0.601 & 11.7 \\
\ce{NbB2} & 191 & 3 & 0 & 394.7 & 0.625 & 11.5 \\
\ce{ReTc} & 187 & 2 & 0 & 199.9 & 0.813 & 11.5 \\
\ce{Ta} & 229 & 1 & 0 & 143.3 & 0.961 & 11.2 \\
\ce{HfN} & 225 & 2 & 0 & 301.8 & 0.690 & 11.1 \\
\end{tabular*}
\end{table}

As discussed in Sec.~\ref{sec:strategy} the training dataset DSI is used to build machine learning models and predict new superconductors. The most interesting predictions (following the criteria detailed in Sec.~\ref{sec:strategy}) are validated from first principles and collected into the validation dataset DSV. 

In this section, we explore all compounds in DBI and DBV that exhibit exceptionally high values of \Tc. For our analysis of specific materials, we rely on calculations conducted using the accurate setting (see Sec.~\ref{sec:methods}), and we determine \Tc\ using the isotropic Eliashberg equation. It is worth noting that the accurate calculation involves substantial computational effort. 
In addition we have chosen three cases of extreme interest to be analysed by a cutting-edge SCDFT methodology, including anisotropy and ab initio Coulomb interactions.   

Some of the compounds in DBI, such as \ce{KB6}~\cite{Katsura2010}, \ce{NbC}~\cite{Gusev1989}, \ce{MgB2}~\cite{Nagamatsu_MgB2_Nature2021}, \ce{NbB2} and \ce{TaB2}~\cite{Buzea2001}, \ce{RuO2}~\cite{Choudhary2022}, etc., are known superconductors or have already been proposed in the literature as superconductors. 
On the other hand, some of the materials in DBV, once computed with a more accurate setting, turn out to be unstable or non-superconducting.
A selection of the most promising compounds is listed in \cref{tab:accuratea}. This table includes a set of materials that our methodology has predicted that could be synthesizable valuable superconductors. 
In the sections below we discuss these materials grouping them into families of compounds. Further information concerning their electronic and phononic band structures and the electron-phonon coupling is given in the SI. 

\subsection{Nitrides}

\Cref{tab:accuratea} contains a large number of nitride compounds with high-\Tc. Above 20~K we find \ce{LiMoN2} (at an incredible $\Tce=46$~K), \ce{ScMoN2} ($\Tce=30$~K), and \ce{LiTcN2} ($\Tce=28$~K). Due to the relevance of this finding, \ce{LiMoN2} is studied in detail in the following. Additionally, we have identified a series of systems with similar chemical compositions that crystallize in the $\gamma$-\ce{LiBO2} (chalcopyrite) structure. Several nitrides adopting this structure have been synthesized in the past, including \ce{MgGeN2}\cite{vanNostrand2005}, \ce{CaGeN2}\cite{Maunaye1971}, \ce{LiPN2}\cite{Marchand1982}, and \ce{NaPN2}\cite{NaPN2}. Furthermore, other A$^\text{II}$B$^\text{IV}$N$_2$ compounds have been produced in the related $\beta$-\ce{NaFeO2} phase, such as \ce{BeSiN2}\cite{Eckerlin1967}, \ce{MgSiN2}\cite{David1970}, \ce{MgGeN2}\cite{David1970}, and \ce{ZnGeN2}\cite{Wintenberger1973}.

The nitride chalcopyrite compound with the highest transition temperature that we have discovered is \ce{LiTcN2}. While this compound is thermodynamically stable, its practical applications are limited due to the presence of the radioactive chemical element Tc. Interestingly, the Fermi energy resides in a valley of the density of states, suggesting that \Tc\ can be enhanced through doping with either electrons or holes. The electron-phonon coupling constant $\lambda$ is calculated to be 1.33, with the majority of the contribution stemming from the acoustic and first optical phonons below 24.8~meV. With an electronic logarithmic average $\olog$ of 241~K, this results in a critical temperature of $\Tce=28$~K.

It is worth mentioning that \ce{LiMoN2} is also dynamically stable in the chalcopyrite structure, with an energy only 56~meV/atom above the convex hull. However, the superconducting critical temperature decreases compared to the trigonal phase. For \ce{LiMoN2} in the chalcopyrite structure, we find $\Tce=22$~K, with an electron-phonon coupling constant of $\lambda=0.99$ and an electronic logarithmic average of $\olog=292$~K.

\subsection{Hydrides}

The remarkable increase in \Tc\ observed in certain hydrogen-rich materials under high-pressure, such as sulfur hydride (\ce{H3S})~\cite{DrozdovEremets_SH3_Nature2015} and lanthanum hydride (\ce{LaH10})~\cite{Eremets_LaH_Nature2019}, has attracted significant attention. These materials can exhibit \Tc\ values soaring as high as 250~K, rivaling or even surpassing those of unconventional superconductors. This remarkable achievement has sparked enthusiasm within the scientific community. However, it is important to note that the practical application of these materials is hindered by the exceedingly high pressures (over 100--200~GPa) required to stabilize them. Our list of compounds with high-\Tc\ at ambient pressure includes several hydride and hydrogen-containing materials. 

The compound with the highest transition temperature among them is \ce{LiPdH2}. In the Li-Pd-H ternary phase diagrams, we have experimental knowledge of \ce{LiPdH}~\cite{Norus1990} (a compound that is not superconducting above 4~K~\cite{Norus1990}), and \ce{Li2PdH2}~\cite{Kadir1989}. Regarding \ce{LiPdH2}, we find that it has a thermodynamically stable tetragonal phase. However, the compound with a high \Tc\ is a trigonal allotrope that crystallizes in the delafossite structure, slightly higher in energy by 101~meV/atom with respect to the tetragonal phase.

The electron-phonon coupling in this compound primarily arises from interactions with high-energy optical phonons, predominantly falling within the range of 75 to 100~meV. The presence of these high-energy phonons is facilitated by the light mass of hydrogen. Consequently, the compound exhibits a significant electron-phonon coupling constant of $\lambda=0.98$ and an exceptionally high $\olog=532$~K. These characteristics contribute to a high critical temperature of $\Tce=43$~K.

The next hydride on our list is \ce{PdH}, that exhibits a critical temperature of $\Tce=37$ K. Palladium hydride is a well-known compound due to its anomalous isotope effect~\cite{Stritzker1972,Schirber1974,Schirber1984}. This effect manifests as an increase in \Tc\ when hydrogen is replaced by deuterium or tritium. The anomalous behavior of \ce{PdH} has been attributed to the presence of strong anharmonic effects~\cite{karakozov1978influence}, which also lead to a substantial reduction in the predicted \Tc\ values obtained using the harmonic approximation (utilized in this study) to approximately 5~K~\cite{Errea2013}. 

\ce{ZrH3} is a hypothetical compound situated 94~meV/atom above the convex hull, indicating its thermodynamic propensity to decompose into \ce{ZrH2} and \ce{H2}. The compound discovered within our framework displays a tetragonal crystal structure, with zirconium atoms occupying the center of a cuboctahedron, while hydrogen atoms are located at the vertices. An intriguing feature of this material is that the \dosef\ is predominantly derived from the Zr states and these electrons are coupled with all phonon modes. Interestingly, a substantial contribution to the electron-phonon coupling constant of $\lambda=0.97$ arises from the acoustic modes, which are exclusively associated with Zr vibrations, leading to a $\Tce=24.9$~K.

\begin{figure*}
\includegraphics[width=0.14\textwidth]{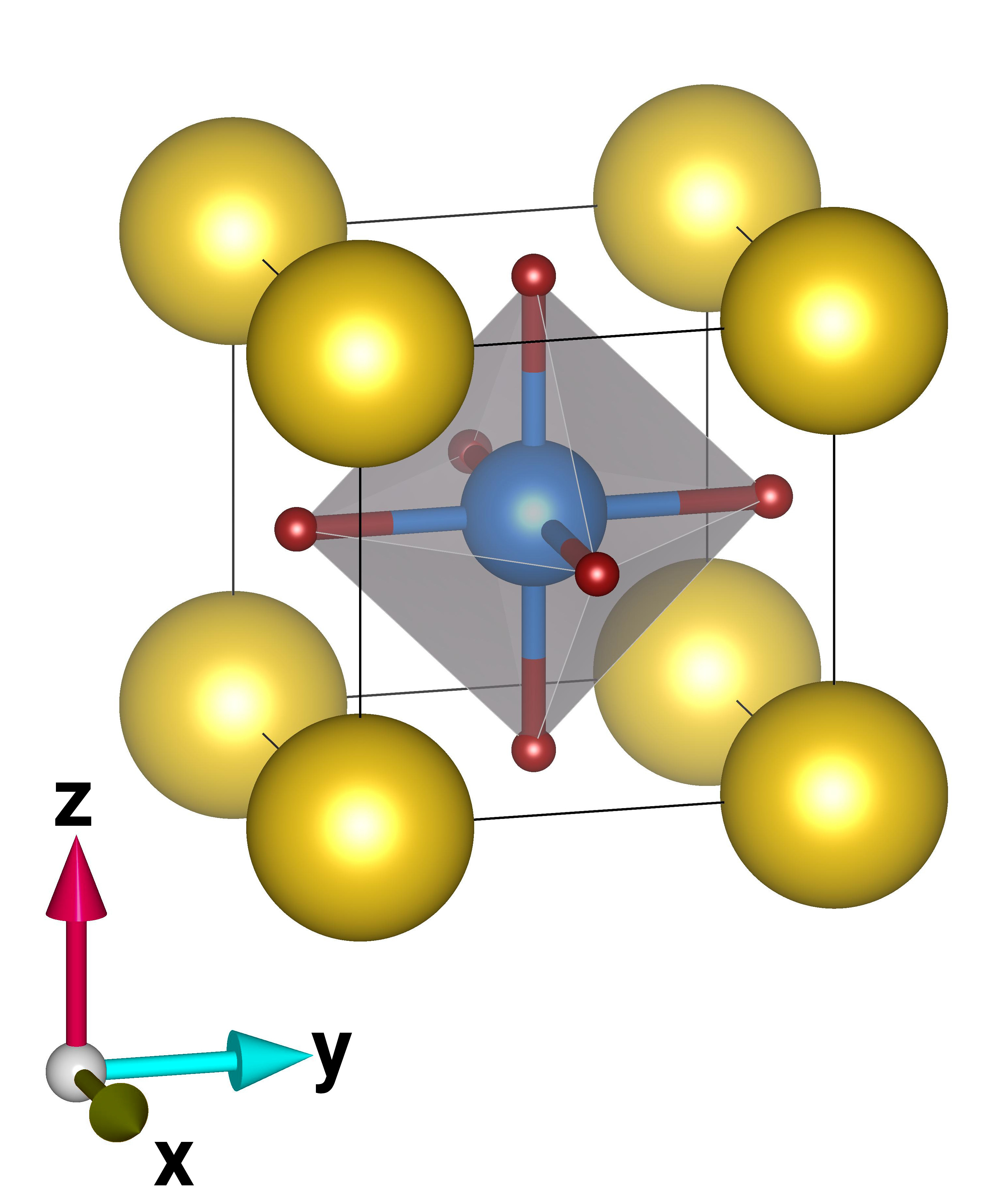} 
\hspace{0.05\textwidth}
\includegraphics[width=0.77\textwidth]{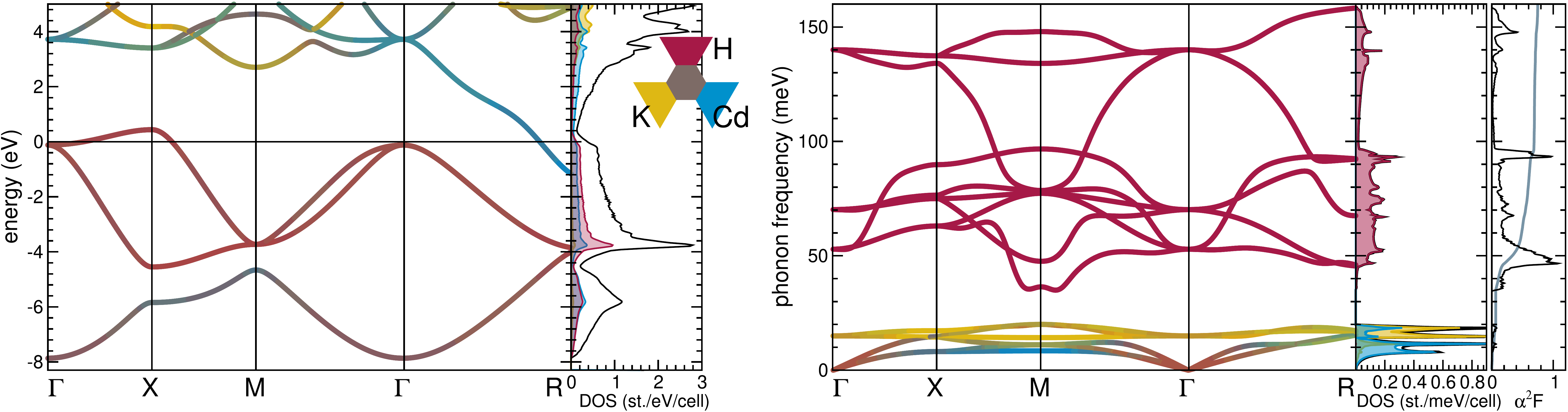} 
\caption{Left: View of the crystal structure of \ce{KCdH3}. The unit cell is shown in black, and the three lattice vectors are shown as arrows. K atoms are in orange, Cd atoms are in blue and H atoms are in red. Center: Atom resolved electronic band structure and density of states which, in the selected energy window, are dominated by Ti states. Right: Atom resolved phonon band structure and density of states. In the rightmost panel are shown the Eliashberg function \afo\ and the integration curve of the electron-phonon coupling $\lambda(\omega)$.}
\label{img:KCdH3}
\end{figure*}

\begin{figure}
 a) KCdH$_{\text{3}}$,\, Tc=23.4K \hspace{0.13\columnwidth} b) Ti$_{\text{3}}$Te,\, Tc=14.8K
 \begin{center} \vspace{-0.3cm}
 \noindent
\includegraphics[width=0.48\columnwidth]{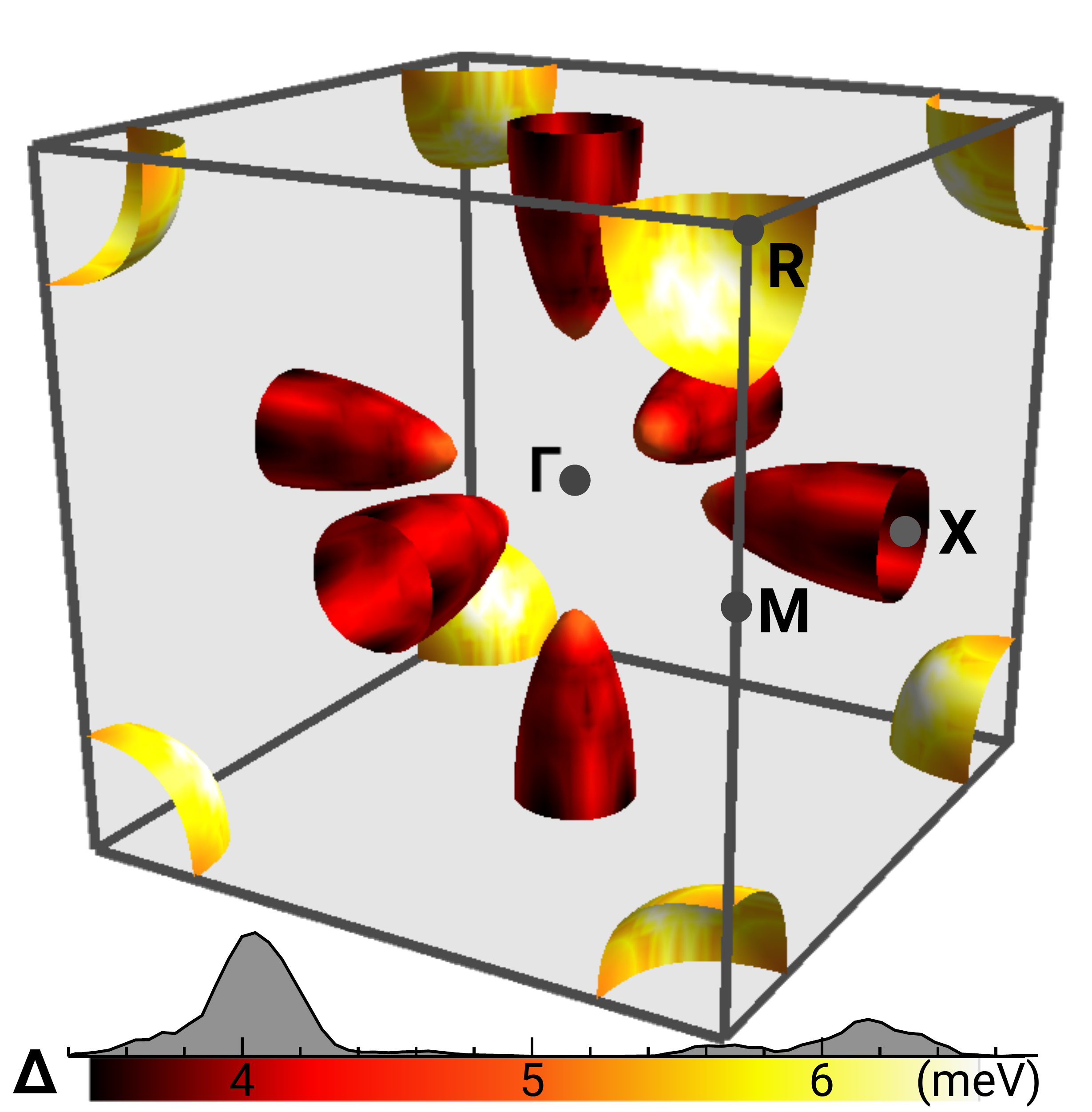} 
\includegraphics[width=0.48\columnwidth]{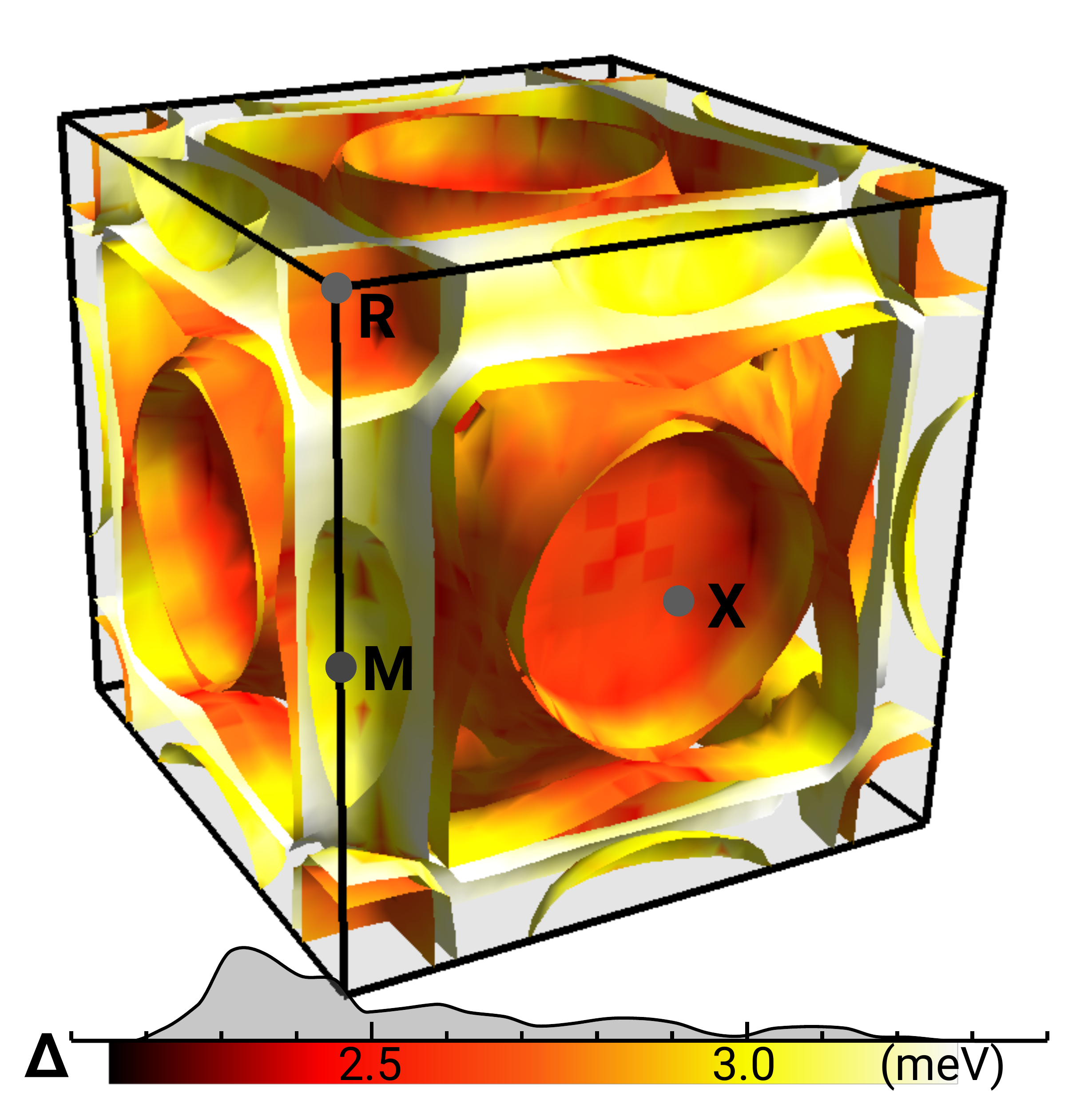} \\
c) LiMoN$_{\text{2}}$, Tc=38K \\\vspace{-0.2cm}
\includegraphics[width=0.48\columnwidth]{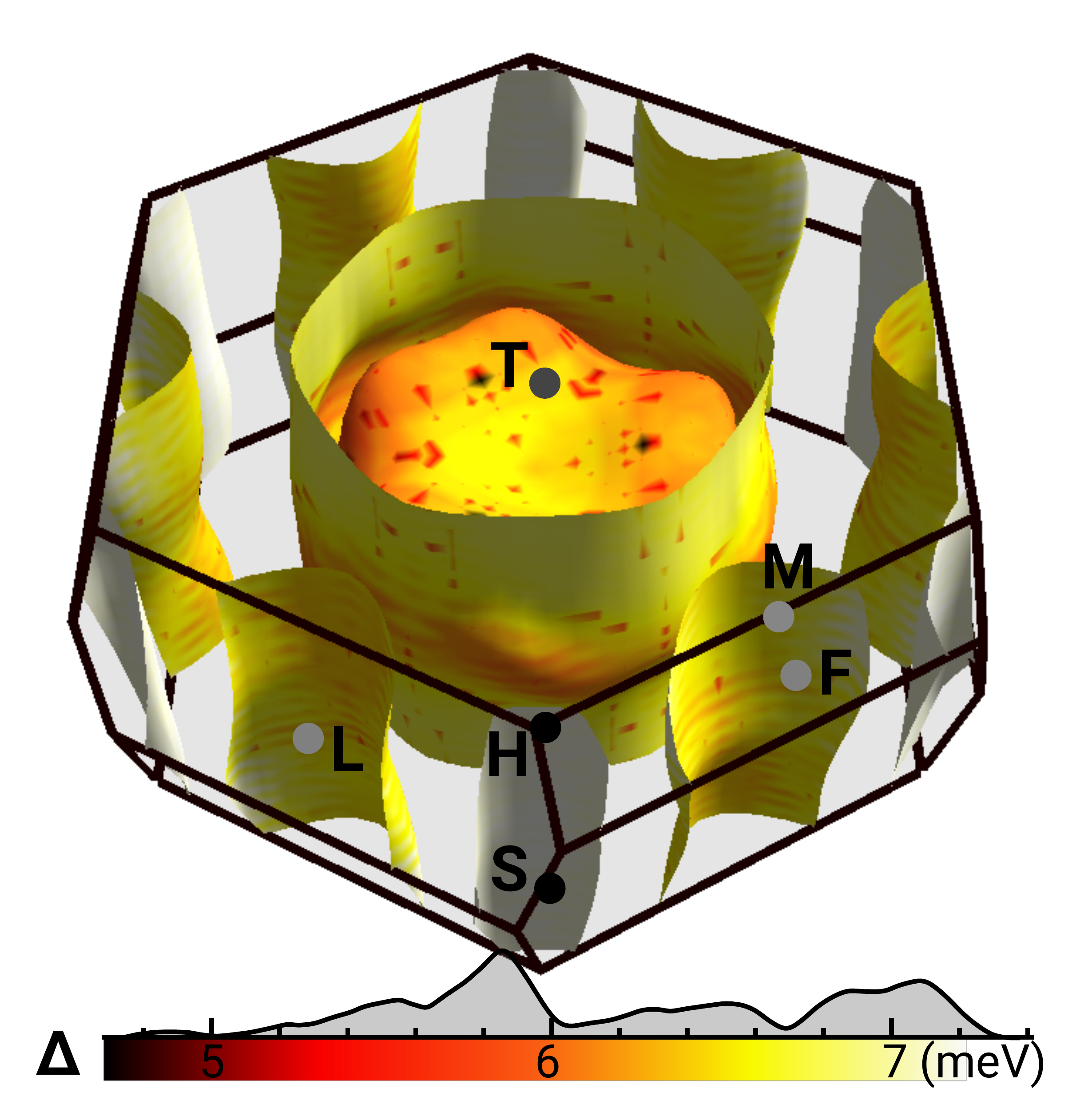}
\caption{Fermi surfaces and gap distributions for three selected compounds. The coloring of the Fermi surface indicates the superconducting gap as shown in the colorbar. The gray curves are gap distribution functions, defined as histograms of the superconducting gap at the Fermi energy. Calculations are performed at high accuracy using fully anisotropic SCDFT.}
\label{fig:FS}
\end{center}
\end{figure}

Finally, we would like to highlight \ce{KCdH3}, a hydride perovskite compound (see \cref{img:KCdH3}). In this case, the bottom of the conduction band falls below the Fermi level at the R point of the Brillouin zone, while the top conduction band is slightly above the Fermi level at the X point. Consequently, the Fermi surface exhibits both electron and hole pockets, with the Fermi level residing in a steep shoulder of the density of states.
This material appears to have an interesting superconducting state, so we decided to study it with the fully anisotropic SCDFT approach (see sect.~\ref{sec:methods}). The hole pocket provides a Fermi surface sheet (see \cref{fig:FS}) located near the X point and has mixed Cd-H character, carrying about 75\% of the total DOS. The nearly spherical electron pocket arises from a pure Cd band and is located at the R-point contributing to about 25\% of the DOS. However both Fermi surface pockets are quite small therefore the intraband electron phonon coupling only comes from low $q$ phonons. These have weak matrix elements and can not sustain superconductivity. Stronger coupling comes only from large $q$ phonons arising via interband scattering. Clearly, owing to the different DOS, the interband scattering mechanism is particularly advantageous for the low DOS electron band which has a coupling about twice that of the hole pocket. It is crucial to note that this points to the fact that \Tc\ is highly sensitive to the precise position of the Fermi level. In our most accurate calculations using SCDFT we estimate a critical temperature of 23.4~K, twice as large as the Eliashberg calculations in \cref{tab:accuratea}. The reason for this discrepancy is not only due to critical convergence aspects but also to the role of superconducting anisotropy which raises \Tc\ by about 5K. 

We should emphasize that anharmonic effects can significantly influence the properties of all hydrides discussed within this section. In order to gain a comprehensive understanding of these effects, further investigations are necessary, such as employing advanced theoretical methods like the stochastic self-consistent harmonic approximation (SSCHA)~\cite{Errea2013,Monacelli2021}.

\subsection{Intermetallics}

We find a wealth of intermetallic materials at the top of our list. Among these, a significant number belong to well-known families of superconducting compounds, such as the A15 and C15 families, or are ternary generalizations of them.

Until the advent of the high-\Tc\ ceramics in 1986, the superconductor with the highest known \Tc\ belonged to the A15 family (\ce{Nb3Ge} with a \Tc\ of 23.2~K). Although the binary phases of this family have been extensively investigated, the chemical space of ternary compounds is relatively unexplored. A15 compounds are cubic with the \ce{Cr3Si} structure type. There are several possibilities to generate ternary phases based on this prototype. For example, by varying the chemical species in the \textit{2a} Wickoff positions one can generate materials such as \ce{Nb6AlSi}~\cite{muller1977supraleitung}, \ce{AlSiMo6}~\cite{Brukl1961}, \ce{V6GeOs}~\cite{Pavlyuchenko1978}, etc. It is also possible to populate the \textit{6c} positions with two different kinds of atoms as in \ce{V2FeGe}~\cite{Kanematsu1986} or in \ce{(NbV)3Si2}~\cite{muller1977supraleitung}.

In \cref{tab:accuratea} we find several A15 compounds, such as \ce{Nb3Zn} with a $\Tc=24.7$~K, \ce{Cr3Os} with a $\Tc=19.1$~K, or \ce{Ti3Te} with a $\Tc=16.3$~K. We also find \ce{V6CoSi} that is a ternary A15 variant with a $\Tc=16.2$~K, and many more of those systems can be found in Table~I in the SI. It is worth noting that we did not discover any high-$\Tc$ systems adopting the structure of \ce{V2FeGe} or \ce{(NbV)3S2}.

\begin{figure*}
\includegraphics[width=0.21\textwidth]{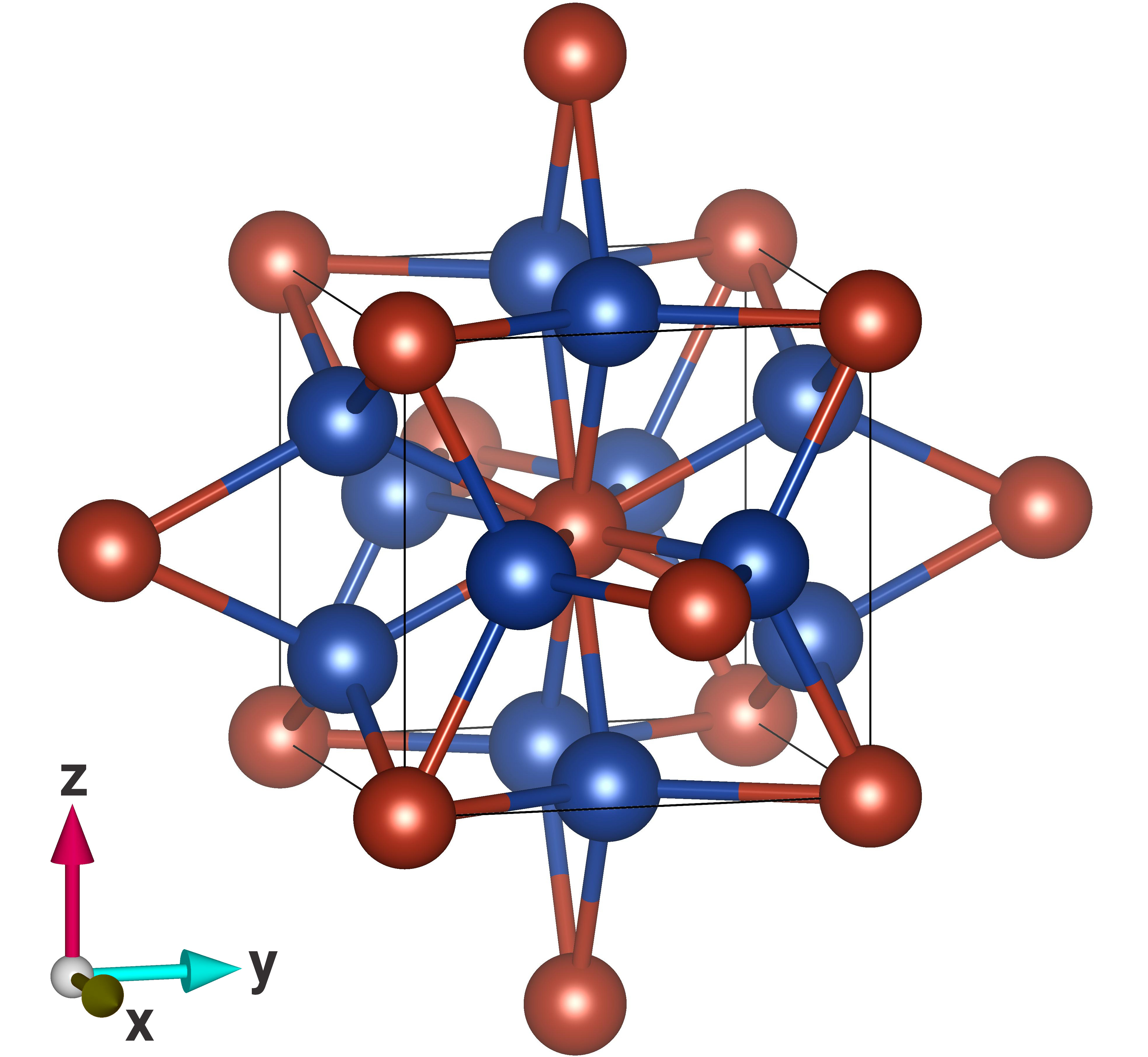} 
\includegraphics[width=0.77\textwidth]{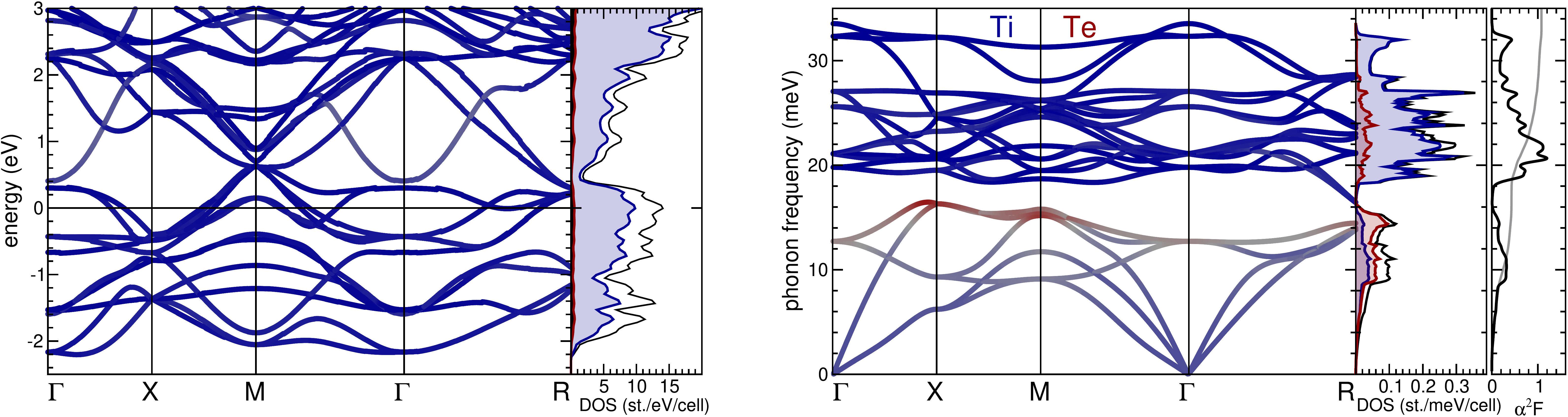} 
\caption{Left: View of the crystal structure of \ce{Ti3Te}. The unit cell is shown in black, and the three lattice vectors are shown as arrows. Ti atoms are in blue, Te atoms are in red. Center: Atom resolved electronic band structure and density of states which, in the selected energy window, are dominated by Ti states. Right: Atom resolved phonon band structure and density of states. In the rightmost panel are shown the Eliashberg function \afo\ and the integration curve of the electron-phonon coupling $\lambda(\omega)$.}
\label{img:Ti3Te}
\end{figure*}

We discuss in more detail \ce{Ti3Te} (see \cref{img:Ti3Te}) due to its unconventional chemical composition, specifically the inclusion of Te, a chemical element rarely found in high-$\Tc$ conventional superconductors. In this compound, the Fermi level occupies a relatively flat region of the DOS, predominantly composed of Ti states. The acoustic modes and the three lowest optical modes extend up to approximately 15.5~meV, with a significant hybridization between the vibrations of Ti and Te. As expected, the higher energy optical states primarily involve the lighter element, Ti. All these modes contribute significantly to the value of $\lambda=1.0$, which, combined with $\olog=187$~K, results in a $\Tc=14.8$~K, as estimated from  SCDFT. The estimation is not far from the 16.3~K obtained with isotropic Eliashberg theory and $\mu^*$=0.10. However the SCDFT estimation accounts both for the anisotropy of the electron phonon coupling, which actually increases \Tc\  by about 2~K and uses the ab-initio calculated electron electron interaction which, as is often the case in transition metals, is quite strong ($\mu$=0.4 and $\mu^*\simeq$ 0.15). The superconducting gap on the Fermi surface is shown in Fig.~\ref{fig:FS}b. The gap distribution is peaked at a value of 2.4~meV featuring a large tail which reaches 3.1~meV on the cylindrical Fermi surface around the RM symmetry line.  

The binary Laves phases, characterized by a cubic C15 structure like \ce{V2Hf}, exhibit a moderate \Tc\ of approximately 10~K. Ternary variations~\cite{Gulay2021} can be introduced through various approaches, resulting in compounds such as \ce{Mg2SiNi3} (possessing trigonal symmetry) or \ce{MgCu4Sn} (exhibiting cubic symmetry). Some of these compounds have been found to superconducting, such  as \ce{Mg2SiIr3} that displays a \Tc\ of 7~K, or \ce{Li2Si3Ir} that exhibits a \Tc\ of 3.8~K~\cite{Kudo2020}.

At the top of \cref{tab:accuratea} with the cubic C15 structure we find \ce{ZrTc2} with a $\Tc=20.2$~K, and a couple of ternary variants of this structure such as \ce{Cr4ReW} with a $\Tc=24.3$~K, \ce{TiNbV4} with a $\Tc=17.7$~K, or \ce{TiV4Mo} with a $\Tc=8.6$~K. Many more systems of this family can be found in Table~I in the SI. As an example, at the Fermi level of \ce{TiV4Mo} we find states with mainly V character and much smaller Ti and Mo character. The Fermi level in this compound is on a shoulder of the DOS (while in \ce{TiNbV4} it is essentially at the maximum of the peak). All phonon modes contribute to $\lambda$, with the largest contribution coming from the optical modes between 12.5 and 30.0~meV. This leads to a $\olog=217.9$~K and $\lambda=0.68$, yielding $\Tc=8.6$~K.

Finally we would like to refer to a series of Ti--V compounds with rather high transition temperatures such as \ce{TiV2} ($\Tc=24$~K), \ce{TiV} ($\Tc=24$~K), and even the elementary substance \ce{V} ($\Tc=24.9$~K). Unfortunately, these transition temperatures are certainly too high due to the neglect of spin fluctuations that are well-known to play an important role in Ti--V compounds~\cite{Matin2014a,Matin2014b,Bose2008},

\subsection{LiMoN$_2$}
\label{sec:limon}

\begin{figure*}
\includegraphics[width=0.22\textwidth]{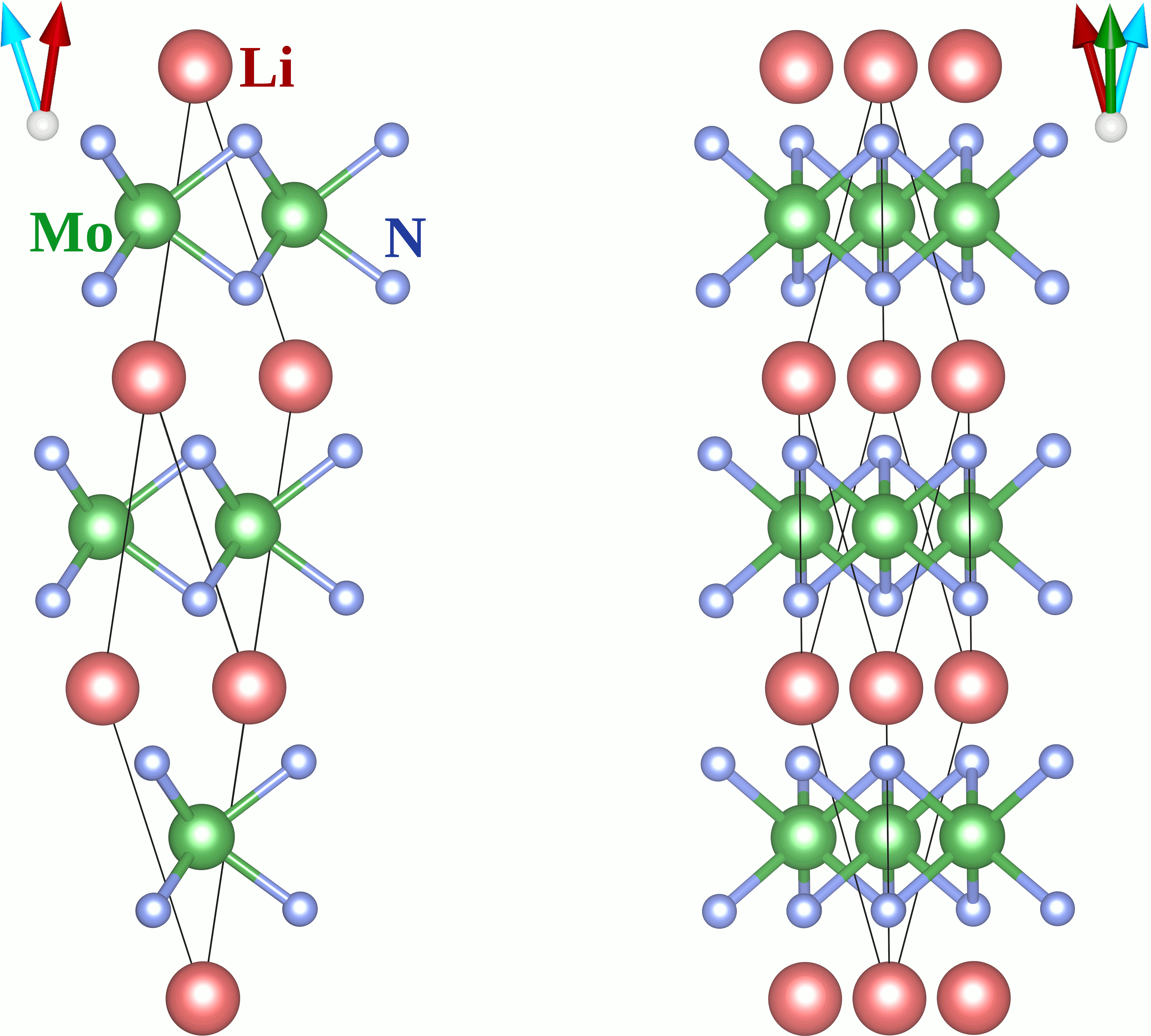} 
\includegraphics[height=0.2\textwidth]{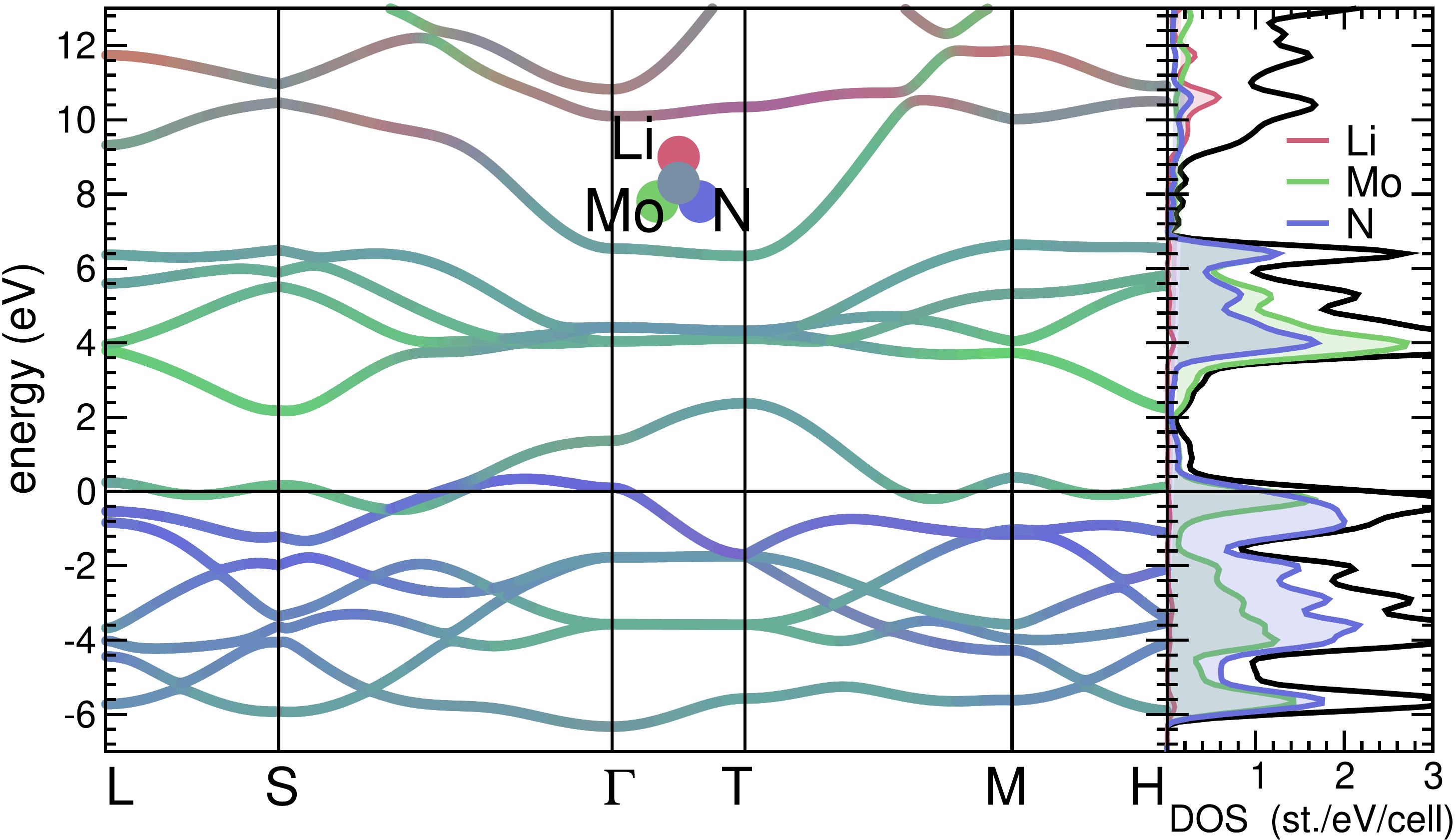} 
\includegraphics[height=0.2\textwidth]{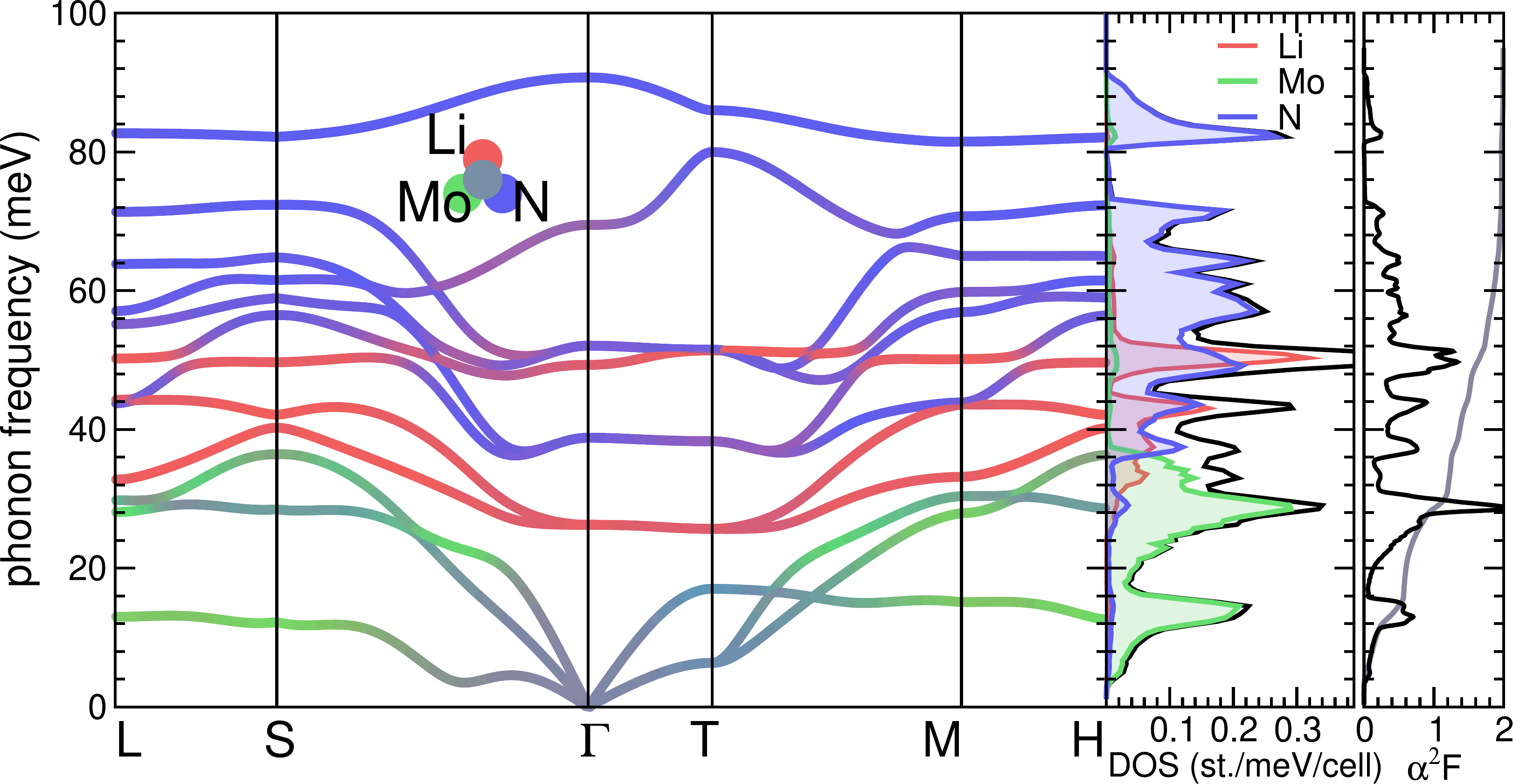} 
\caption{Left: Two views of the crystal structure of \ce{LiMoN2}. The unit cell is shown in black, and the three lattice vectors are shown as arrows. Li atoms are in red, Mo atoms are in green and N atoms in blue. This structure can be pictured as consisting of \ce{MoN2} layers intercalated with Li.  Center: Atom resolved electronic band structure and density of states. Right: Atom resolved phonon band structure and density of states. In the rightmost panel are shown the Eliashberg function \afo\ and the integration curve of the electron-phonon coupling $\lambda(\omega)$.  }
\label{img:LiMoN2}
\end{figure*}

\ce{LiMoN2} is a material that was overlooked in the search for supercondutors. It was synthesized for the first time in 1992 by Elder and coauthors~\cite{elder1992lithium}, that were looking for nitride counterparts to the high-\Tc\ ceramics, since N$^{-III}$ is the ion most similar to O$^{-II}$ with respect to size, polarizability, and electronegativity. This compound is a member of a family of layered nitrides that also includes \ce{MgMoN2}~\cite{verrelli2017viability},
\ce{MnMoN2} and \ce{FeWN2}~\cite{bem1996synthesis}, 
\ce{MnWN2}, \ce{NiWN2} and \ce{CoWN2}\cite{herle1995synthesis},
\ce{LiWN2}~\cite{herle1994synthesis},
\ce{CaTaN2}\cite{balbarin1996high},
\ce{CaNbN2}\cite{CaNbN2_1},
\ce{NaTaN2} and \ce{NaNbN2}~\cite{niewa1996group}, 
\ce{SrZrN2} and \ce{SrHfN2}~\cite{gregory1996synthesis},
\ce{SrTiN2}~\cite{farault2003crystal},
\ce{CuNbN2}~\cite{zakutayev2014experimental},
\ce{LiSrGaN2}~\cite{park2003synthesis}, \ce{CrWN2}~\cite{weil1997chemical},
\ce{CuTaN2}\cite{zachwieja1991cutan},
\ce{BaZrN2}~\cite{seeger1994synthesis},
\ce{BaHfN2}, \ce{BaZr_{1-x}Hf_{x}N2}~\cite{gregory1998synthesis}. From these, \ce{CaTaN2} and \ce{CaNbN2} are known to be superconductors, with \Tc\ of $\sim 8.2$~K and $\sim 14$~K, respectively~\cite{balbarin1996high,oliva2000electronic}. Note that these two compounds were also found by our methodology.

The structure of \ce{LiMoN2} is depicted in the left panel of Fig.~\ref{img:LiMoN2} and can be conceived as being composed of \ce{MoN2} layers intercalated by Li. Interestingly, the \ce{MoN2} layers are isostructural to the two-dimensional transition-metal dichalcogenides, and can be obtained by replacing S$\rightarrow$N in \ce{MoS2}. This leads, of course, to a charge destabilization, that is compensated by the extra electron from Li, leading to a metallic state. Of course, intercalating a divalent atom such as Mg leads to a semiconducting ground-state~\cite{verrelli2017viability} where the two electrons fill the band formed by the lowest d$_{z^2}$ orbitals~\cite{Mattheiss1973energy}.

It is true that are striking structural similarities between \ce{LiMoN2} and other intercalated compounds such as graphite (that when intercalated with Ca reaches transition temperatures above 10~K~\cite{Weller2005,Smith2015superconductivity}). However, a closer look reveals that the separation between the opposing N sheets is smaller than the in-plane separation, suggesting that there is substantial direct bonding between opposing N layers~\cite{singh1992electronic}. Furthermore, an analysis of the electronic structure shows that \ce{LiMoN2} is in fact a three-dimensional metal~\cite{singh1992electronic,oliva2000electronic}. As a two-dimensional band structure (characteristic of the high-\Tc\ ceramics or of \ce{MgB2}) was seen as an essential requirement for high-temperature superconductivity, \ce{LiMoN2} was quickly dismissed after the initial synthesis as not being interesting~\cite{oliva2000electronic}.

The electronic band-structure of \ce{LiMoN2} is depicted in the center panel of Fig.~\ref{img:LiMoN2}, together with the atom-resolved and total density of electronic states. We will keep the description of the electronic properties of \ce{LiMoN2} brief, as they  have already been discussed in Refs.~\onlinecite{singh1992electronic,oliva2000electronic}. There are two bands crossing the Fermi level that are composed of d-states of Mo strongly hybridized with p-states of N. There is a large dispersion of the bands in the $\Gamma$--T direction, perpendicular to the basal planes, indicating the three-dimensional character of the metal. We can see virtually no contribution of Li to the density-of-states indicating that these atoms are fully ionized. Finally, the Fermi level lies on a very large peak of the density-of-states, a fact that is often associated to superconductivity.

The phonon dispersion is plotted on the right panel of Fig.~\ref{img:LiMoN2}. The material is stable dynamically, with three very different sound velocities showing the marked anisotropy of the compound. The three acoustical branches are essentially composed of Mo vibrations, as expected by the larger mass of Mo with respect to Li and N. The Li vibrations contribute mostly to the optical branches between 25 and 42~meV and have a dispersionless Einstein mode at 50~meV. The high-lying phonon branches until 95~meV are to a large extent composed of N vibrations. This is somewhat surprising in view of the mass difference between Li and N, and attests to the strength of covalent nitrogen bonds in the structure. Along the S-$\Gamma$ line we observe an incipient phonon instability corresponding to an in plane charge density wave distortion (CDW). An accurate analysis shows that the frequency of this CDW mode is numerically very unstable, with the mode softening at low electronic temperatures. The phonon frequencies remain, however, real in our most precise calculations.

Also in Fig.~\ref{img:LiMoN2} we show the Eliashberg function $\alpha^2F(\omega)$ and electron-phonon coupling constant $\lambda(\omega)$ of \ce{LiMoN2}. We see that almost all phonon branches (with the exception of the last ones) interact very strongly with the electronic states of Mo and N. As such the plot for $\alpha^2F(\omega)$ follows the features of the Mo and N phonon density-of-states until around 55~meV. Obviously, the Li vibrations do not couple to the electrons, as the bands at the Fermi level do not have any Li s-character. The final value of the electron-phonon coupling constant is $\lambda=1.95$.  This is a very large value, on par with the best conventional superconductors known to date which are the high pressure hydrides (for which $\lambda\sim2$). As a comparison \ce{MgB2}, the best phononic room pressure superconductor, has a coupling of $\lambda\sim0.7$.

One should observe that the Fermi level is placed near a steep peak in the density of states, implying that the large coupling is, in part, caused by a large density of electronic states. Therefore the value of $\lambda$ can be misleading, as superconductivity arises from a broad region with a width of many hundredths of electron volts. Moreover the coupling is unimportant unless associated with stiff phonon modes. In fact, large values of $\lambda$ are often associated with soft-modes and lattice instabilities which do not necessarily lead to superconductivity. As seen in Fig.~\ref{img:LiMoN2} this is not the case for \ce{LiMoN2} where the logarithmic average phonon frequency is $\sim25$~meV, which is small when compared with the hydrides (where it can be higher than 100~meV) and also smaller than for \ce{MgB2} (about 60~meV), but still quite large as compared to most non-hydrogenic superconductors (13~meV in \ce{NbSe2}, 12~meV in \ce{Nb3Sn}, 6~meV in \ce{Pb}, 16~meV in \ce{Nb}, for example).   

We predict a quite high critical temperature of approximately 40~K, similar to that of \ce{MgB2}. This value is significantly smaller than one would expect from a standard McMillan approach with $\mu^{*}=0.10$, that predicts $\Tc=52$~K. The reason is in part to be ascribed to the strong variation of the density of states at the Fermi level, which causes a rapid drop of the electron phonon coupling and, most importantly, a very strong Coulomb repulsion. The latter, similarly to the electron phonon coupling, is increased by the high density of states at the Fermi level. However, it is weakly renormalized as the system has almost no states available between 1 and 3.5~eV above the Fermi level. A very large $\mu^*=0.18$ would be needed, in an McMillan approach, to account for such effects.

We should point out that a Fermi energy located at a peak in the density of states suggests a possible incipient instability of the system, and most importantly also implies that the predicted value of \Tc\ can depend considerably on the calculation parameters and approximations. For example, changing the functional to the standard Perdew-Burke-Ernzerhof approximation~\cite{PBE}, increases slightly \dosef, and causes a phonon softening, leading to an increase of \Tc\ by almost 50\%.

At zero temperature, the energy gap has values between 5 and 7~meV, with three gap regions. Anisotropy is remarkable but superconductivity is largely sustained by inter-band  electron-phonon scattering, unlike the case of \ce{MgB2} where coupling mostly occurs within B$\sigma$ orbitals. This leads to the fact that the three gaps have similar values. Nevertheless anisotropy has an impact on \Tc: calculations done assuming an isotropic coupling yield a slightly smaller \Tc\ of 35~K. 

An important question is if the high value of \Tc\ predicted here can be realized experimentally. In our opinion this is mainly dependent on the quality of the experimental samples. The major problem seems to be related to disorder and incipient instabilities. The original synthesis resulted in a 15\% concentration of Li$_\text{Mo}$ anti-sites~\cite{elder1992lithium}. This can be understood as both Li and Mo sites are sixfold coordinated with N and that the cation--N distances are similar~\cite{singh1992electronic}. One can argue~\cite{singh1992electronic} that disorder would decrease \dosef, with a negative impact on superconductivity. On the other hand, it was found that up to 64\% of the Li could be deintercalated from \ce{LiMoN2}~\cite{elder1992lithium}. This could be used to lower the position of the Fermi level, increasing the density-of-states at the Fermi level, and, in a first approximation, increasing the superconducting transition temperature. Furthermore, the fact that \ce{LiMoN2} is close to a charge-density state can lead to structural deformations and consequent decrease of \dosef. Other problems could arise from the difficult in forming single-phase \ce{LiMoN2} and the presence of secondary phases~\cite{hunter2008structural}.

\section{Conclusions}
\label{sec:conclusion}

We conducted an extensive investigation into conventional superconductivity by combining state-of-the-art calculations of the electron-phonon coupling with machine learning-accelerated high-throughput techniques. To achieve this, we created a comprehensive dataset comprising over 8250 ab-initio electron-phonon calculations. This dataset represents a significant leap forward, as it is at least one order of magnitude larger than any previously available computational dataset for conventional superconductors. Furthermore, our calculations exhibit a good level of convergence, allowing us to identify intriguing superconducting materials and thoroughly examine the electron-phonon and superconducting properties across the entire spectrum of stable compounds.

Our dataset served as a foundation for training a machine learning model, leveraging compositional, structural, and ground-state properties as input features. Equipped with this powerful machine learning model, we explored a materials space encompassing approximately 200\,000 metallic compounds. Our goal was to identify all superconducting compounds predicted to possess a $\Tc$ greater than 5~K. The model achieved a 65\% success rate in this task. For comparison, in our initial dataset, only 10\% of the materials fell within this range. Considering the full set of materials with calculated $\Tc$ greater than 5~K studied, 55\% of these were suggested by the machine learning model (60\% if we consider materials with $\Tc\ > 10$~K), which attests the effectiveness of our approach.

Regarding the behavior observed across the materials space, we discovered that both \olog\ and $\lambda$ appear to follow a lognormal distribution. This distribution is asymmetric as these quantities cannot be negative. However, the underlying reasons for this distribution pattern remain currently unknown. This knowledge enabled us to estimate that the likelihood of encountering a metal with a $\lambda$ value exceeding 1, indicating a very strong electron-phonon coupling, is approximately 1\%. Consequently, such compounds are rare occurrences within the materials space. Assuming uncorrelated lognormal distributions for \olog\ and $\lambda$, we observe a superexponential distribution of \Tc\ values. 
However the presence of large $\lambda$ values often arises from soft phonons which lead to small \olog\ values  truncating the tail of the \Tc\ distribution. In terms of probability to find superconductivity, our analysis suggests that within the set of stable, metallic and non-magnetic materials, there is around a 0.4\% chance of $\Tc > 20$~K, and a 0.03\% chance of superconductivity above 30~K. This finding aligns well with the long-held prejudice that conventional superconductivity is limited to approximately this temperature range. What we quantify here is that high \Tc\ superconductivity with the phononic mechanism is so rare, in stable compounds, that an unbiased search is hopelessly inefficient and acceleration methods are necessary.

We observe a series of regularities in what concerns the chemistry of the materials with higher values of \Tc. When considering metallic elements, Ti and V from the 4th row, as well as Zr, Nb, Mo, Tc, and Ru from the 5th row of the periodic table, display favorable characteristics for superconductivity. Additionally, many intermetallic compounds with high \Tc\ values belong to well-established families, such as the A15 or C15 structures, or are ternary extensions of these. Despite an extensive exploration of numerous systems, we did not discover any compound of this nature surpassing a transition temperature of approximately 24~K. In fact, the most exceptional \Tc\ values were found in compounds containing non-metallic elements, including H, N, O, and others. We identified 73 hydrides with \Tc\ exceeding 5~K, as well as 45 nitrides, 31 carbides, and so on. It is worth emphasizing that our study also unveiled numerous superconducting compounds with unconventional chemical compositions or unique crystal structures.

The compound with highest superconducting transition temperature in our study was the layered metal \ce{LiMoN2} with $\Tc\sim38$~K. This value of \Tc\ can be understood by the extreme electron-phonon coupling between the electrons participating in the very strong covalent bonds within the \ce{MoN2} layers, with almost all N and Mo phonon modes contributing equally to a value of $\lambda=1.9$. The material exhibits three different superconducting gaps, but, contrary to \ce{MgB2}, superconductivity seems to arise mainly from interband coupling. These results show that high-\Tc\ superconductivity can exist in metallic layered nitride compounds and call for a detailed experimental analysis of these materials, and in particular of \ce{LiMoN2}.

The synergy of machine-learning techniques and conventional density-functional based approaches holds great potential for a systematic exploration of the multinary phase diagram, enabling the search for superconducting compounds at high temperatures, and even room temperature. Furthermore, the availability of more data will inevitably make machine learning models more precise, in a virtuous cycle that will allow the community, in a near future, to investigate conventional superconducting properties of all possible stable materials, both at ambient as well as under pressure. 

\section{Methods}
\label{sec:methods}

\subsection{Pseudopotentials}

We use the Perdew-Burke-Ernzerhof for solids~\cite{pbesol} (PBEsol) pseudopotentials from the \textsc{pseudodojo} project~\cite{vanSetten2018pseudodojo}, specifically the stringent, scalar-relativistic norm-conserving set. This pseudopotential table has been systematically constructed and validated in a series of 7 tests in crystalline environments, specifically the $\Delta$-Gauge~\cite{Lejaeghere2013}, $\Delta'$-Gauge~\cite{Jollet2014}, GBRV-FCC, GBRV-BCC, GBRV-compound~\cite{Garrity2014}, ghost-state detection, and phonons at the $\Gamma$-point.

The \textsc{pseudodojo} set includes most chemical elements of the periodic table. Exceptions are lanthanides (although La and Lu are included) and actinides. We noticed severe convergence problems with Ir, that made us replace this pseudopotential by a previous version. For the cutoff energies, we use the maximum of \textsc{pseudodojo}'s high precision hint for the elements in a given material. 

\subsection{Electron-phonon}

All density-functional calculations are performed using the versions 6.8 and 7.1 of \textsc{quantum espresso}~\cite{Giannozzi2009,Giannozzi2017} with the Perdew-Burke-Ernzerhof (PBE) for solids (PBEsol)~\cite{pbesol} generalized gradient approximation.

Geometry optimizations are performed using uniform $\Gamma$-centered $k$-point grids with a density of 1500~$k$-points per reciprocal atom. If this results in an odd number of $k$-points in a given direction, the next even number is used instead. Convergence thresholds for energies, forces and stresses are set to $1\times10^{-8}$\,a.u., $1\times10^{-6}$\,a.u., and $5\times10^{-2}$~kbar, respectively. For the electron-phonon coupling we use a double-grid technique, with the same $k$-grid used in the lattice optimization as the coarse grid, and a $k$-grid quadrupled in each direction as the fine grid. For the $q$-sampling of the phonons we use half of the $k$-point grid described above. The double $\delta$-integration to obtain the Eliashberg function is performed with a Methfessel–Paxton smearing of 0.05~Ry.

For the higher accuracy calculations, we repeat the previous steps by changing: (i) the initial $k$-point grid density is set to 3000~$k$-points per reciprocal atom; (ii) the $k$-grid used as the coarse grid is set to the double of the $k$-grid used for the geometry optimization.

One of the main problems that we encountered is related to imaginary phonon frequencies. There are several aspects of this problem. First, compounds often exhibit imaginary frequencies at $\Gamma$ due to the breaking of translation symmetry caused by numerical imprecision, either due to a too low energy cutoff or too few $k$-points, for example. To circumvent this problem, we accept calculations where we encounter at most 3 imaginary frequencies at $\Gamma$ if the maximum imaginary frequency is below $35i$~cm$^{-1}$. We also encountered a few systems exhibiting spurious soft modes due to insufficient $k$-point sampling leading to instabilities. Finally, the $q-$point sampling may miss some phonon instabilities in undersampled regions of the Brillouin zone. All these problems lead to either false positive or false negative entries in our dataset. The former are eventually detected in the high accuracy step of our workflow. The false negatives, and in particular those compounds that our methodology labels as unstable, are unfortunately overlooked.

\subsection{Superconductivity}
\label{sec:met:super}

The values of
\begin{subequations}
\begin{align}
  \lambda & = 2 \int \frac{\alpha^2F(\omega)}{\omega} \;d\omega, \\
  \log(\omega_\text{log}) & = \int \frac{\log(\omega)}{\omega}\; \alpha^2F(\omega) \;d\omega ,\\
  \omega_2^2 & = \int \omega\; \alpha^2F(\omega) \;d\omega,
\end{align}
\end{subequations}
where $\alpha^2F(\omega)$ is the Eliashberg spectral function, are used to calculate the superconducting transition temperature using the McMillan formula~\cite{Mcmillan1968tc,Dynes1972}
\begin{equation}
    T_\text{c}^\text{McMillan} = \frac{w_\text{log}}{1.20} \exp\left[-1.04\frac{1 + \lambda}{\lambda - \mu^*(1 + 0.62\lambda)}\right]
    \,,
\end{equation}
and the Allen-Dynes modification~\cite{Allen1975} to it:
\begin{equation}
    T_\text{c}^\text{AD} = f_1 f_2 T_\text{c}^\text{McMillan}
    \,,
\end{equation}
where the correction factors are
\begin{subequations}\begin{align}
    f_1 & = \left\{1 + \left[\frac{\lambda}{2.46(1 + 3.8*\mu^*)}\right]^{3/2}\right\}^{1/3} \\
    f_2 & = 1 + \frac{\lambda^2 (\omega_2/\omega_\text{log} - 1)}
    {\lambda^2 + \left[1.82(1 + 6.3\mu^*) \omega_2/\omega_\text{log}\right]^2}
\end{align}\end{subequations}
The function $\alpha^2F(\omega)$ is also used to obtain \Tc\ from the solution of the isotropic Eliashberg equations.

We took arbitrarily the value of $\mu^*=0.10$ for all materials studied. We note that this procedure is well defined for the McMillan's and Allen-Dynes formulas, but not for the Eliashberg equations. Indeed, these depend on an extra parameter, the cutoff of the Coulomb interaction, and for which we took the (rather arbitrary) value of 0.5~eV.

\subsection{DFT for superconductors}

The density of states of \ce{LiMoN2} implies a strong energy variation of the gap, and the band structure indicates the likeliness of superconducting anisotropy (linked to the N and Mo orbital character at the Fermi level). For this reason to study superconductivity we must adopt an anisotropic approach, accounting both for the anisotropy of the Fermi surface and the energy dependence of the electronic states. Presently the only approach that can describe all these physical effects is superconducting density-functional theory~\cite{Lueders_SCDFT_PRB2005,Marques_SCDFT_PRB2005} (SCDFT). In this theory, the only quantity we describe as isotropic is the Coulomb interaction, as tests have shown that its anisotropy has no significant effect in the results. We use the most recent SCDFT functional~\cite{SPG_EliashbergSCDFT_PRL2020}, where the gap equation is solved for the Kohn-Sham gap, while the physical superconducting gap is computed from it as a post processing step.

A strict energy cutoff of 104~Ry is used for \ce{LiMoN2} and phonons and electron phonon couplings were computed  on a $12\times12\times12$ and $6\times6\times6$ grid for $k$- and $q$-points respectively. These are interpolated to a set of 80\,000 $k$-points on the Fermi surface, which is used for the SCDFT simulations~\cite{Sanna_NbSe2_npjQM2022}. For the calculation of the screened Coulomb interaction we have used the RPA approximation for the screening function which was computed on a  $8\times8\times8$ $q$-grid.

The same framework is used to study \ce{KCdH3} and \ce{TiTe3}. For \ce{KCdH3} we converge our results using a 120~Ry cutoff, a $16\times16\times16$ ($8\times8\times8$) grid for $k$ ($q$)-points for the calculation of the phonons and an $8\times8\times8$ $q$-grid for the screening function. For \ce{KCdH3} we use a 120~Ry cutoff, an $8\times8\times8$ ($6\times6\times6$) grid for $k$ ($q$)-points for the calculation of the phonons and a $6\times6\times6$ $q$-grid for the screening function. 

\subsection{Machine Learning}
\label{sec:methMachineLearning}

In order to predict directly the Debye temperature we train a model based on the \textsc{alignn} network~\cite{Choudhary2021}. For the training data, we use the dataset from Ref.~\onlinecite{deJong2015}, containing 10987 entries based on data from the materials project~\cite{materialsproject}. We split the dataset randomly using a 80/10/10 split. Due to the relationship between $\Theta_\text{D}$ and the elastic constants~\cite{anderson1963simplified} we use the same hyperparameters as the best model trained for the bulk and shear modulus present in the MatBench repository~\cite{Dunn2020}. The error obtained by the trained model in the test set is 25.3~K, while predicting the average of the train set (345.7~K) would result in an error of 133.2~K.

Looking at MatBench~\cite{Dunn2020}, we see that for datasets of similar size to our superconducting data, the model that yields better results is \textsc{modnet}~\cite{DeBreuck2021}. We train \textsc{modnet} using as targets, simultaneously, $\lambda$, \olog\ and \Tc\, with the error for each property weighted equally, as these are the choices yielding the best results. Hyperparameters are optimized using a grid search approach and 5-fold cross-validation (see SI for details on the grid search and optimized parameters). For the final model we use the ensemble of the 5 models with smallest cross-validation error.

\section{Data availability statement}

All data used in or resulting from this work will be available in Materials Cloud and in NOMAD as soon as the manuscript is accepted for publication. The workflow and machine-learning models will be released on github.

\section{Acknowledgements}

T.F.T.C acknowledges financial support from FCT - Fundação para a Ciência e Tecnologia, Portugal (projects UIDB/04564/2020 and 2022.09975.PTDC) and the Laboratory for Advanced Computing at University of Coimbra for providing HPC resources that have contributed to the research results reported within this paper. M.A.L.M. acknowledges partial funding from Horizon Europe MSCA Doctoral network grant n.101073486, EUSpecLab, funded by the European Union, and from the Keele Foundation. M.A.L.M. gratefully acknowledges the Gauss Centre for Supercomputing e.V. (\url{www.gauss-centre.eu}) for funding this project by providing computing time on the GCS Supercomputer SuperMUC-NG at Leibniz Supercomputing Centre (\url{www.lrz.de}).

\section{Author Contributions}

T.F.T.C. and M.A.L.M. performed the high-throughput \emph{ab initio} calculations. T.F.T.C trained the machine learning models. A.S. developed the Eliashberg solver and performed the SCDFT calculations. All authors contributed to designing the research, interpreting the results and writing of the manuscript. 

\section{Competing  Interests}

The authors declare that they have no competing interests.

\bibliography{references.bib}

\end{document}